\documentclass[
floatfix,prx,aps,a4paper,showpacs]{revtex4-2}
\pdfoutput=1

\usepackage{graphicx}
\usepackage{amssymb}
\usepackage{amsmath}
\usepackage[utf8]{inputenc}
\usepackage{cancel}
\usepackage[usenames,dvipsnames]{color}
\usepackage{psfrag}
\usepackage{epsfig}
\usepackage{bbm}
\usepackage{bm}
\usepackage{dsfont}
\usepackage{soul}
\usepackage{colonequals}
\usepackage{float}
\usepackage{enumitem}
\definecolor{TURed}{RGB}{153,0,0}
\definecolor{DarkGreen}{RGB}{0,153,0}
\definecolor{DarkBlue}{RGB}{0,0,153}
\newcommand{\nn}{\nonumber\\}

\begin{document}

\title{
Irreversibility, heat and information flows induced by non-reciprocal interactions
} 
\author{Sarah A.~M.~Loos
	and Sabine H.~L.~Klapp}
\affiliation{
  Institut f\"ur Theoretische Physik,
  Hardenbergstr.~36,
  Technische Universit\"at Berlin,
  D-10623 Berlin,
  Germany}
\date{\today}

%%%%%%%%%%%%%%%%%%%%%%%%%%%%%%%%%%%%%%%%%%%%%%%%%%%%%%%%%%%%%%%%%%%%%%%%%%%%%%%%%%%%%%%%%%%%%%%%%%%%%%%%%%%
%%%%%%%%%%%%%%%%%%%%%%%%%%%%%%%%%%%%%%        ABSTRACT        %%%%%%%%%%%%%%%%%%%%%%%%%%%%%%%%%%%%%%%%%%%%%%%%%%%%%%%%%
%%%%%%%%%%%%%%%%%%%%%%%%%%%%%%%%%%%%%%%%%%%%%%%%%%%%%%%%%%%%%%%%%%%%%%%%%%%%%%%%%%%%%%%%%%%%%%%%%%%%%%%%%%%
\begin{abstract}
We study the thermodynamic properties induced by non-reciprocal interactions between stochastic degrees of freedom in time- and space-continuous systems. We show that, under fairly general conditions, non-reciprocal coupling alone implies 
a steady energy flow through the system, i.e., non-equilibrium. Projecting out the non-reciprocally coupled degrees of freedom renders non-Markovian, one-variable Langevin descriptions with complex types of memory, for which we find a generalized second law involving information flow. 
We demonstrate that 
non-reciprocal linear interactions can be used to engineer non-monotonic memory, which is typical for, e.g., time-delayed feedback control, and is automatically accompanied with a nonzero information flow through the system. Furthermore, already a single non-reciprocally coupled degree of freedom can extract energy from a single heat bath (at isothermal conditions), and can thus be viewed as a minimal version of a time-continuous, autonomous ``Maxwell demon". We also show that for appropriate parameter settings, the non-reciprocal system has characteristic features of active matter, such as a positive energy input on the level of the fluctuating trajectories without global particle transport.
 \end{abstract}

\maketitle

%%%%%%%%%%%%%%%%%%%%%%%%%%%%%%%%%%%%%%%%%%%%%%%%%%%%%%%%%%%%%%%%%%%%%%%%%%%%%%%%%%%%%%%%%%%%%%%%%%%%%%%%%%%
%%%%%%%%%%%%%%%%%%%%%%%%%%%%%%%%%%%%%%        INTRO        %%%%%%%%%%%%%%%%%%%%%%%%%%%%%%%%%%%%%%%%%%%%%%%%%%%%%%%%%
%%%%%%%%%%%%%%%%%%%%%%%%%%%%%%%%%%%%%%%%%%%%%%%%%%%%%%%%%%%%%%%%%%%%%%%%%%%%%%%%%%%%%%%%%%%%%%%%%%%%%%%%%%%
\section{Introduction}\label{sec:INTRO}
Fundamental physical interactions between
%Most physical systems consisting of 
mutually coupled particles, such as atoms or molecules, %atoms in a polymer, colloids in a dense suspension, or billiard balls,
%involve 
are typically
\textit{reciprocal}. 
They
are 
derivable from a Hamiltonian (i.e., conservative) and thus fulfill, automatically, Newton's third law, \textit{actio = reactio}.
In the absence of driving forces or (temperature) gradients, systems with reciprocal interactions equilibrate and are well described by 
traditional thermodynamics. This holds even on the mesoscale, that is, when instead of the full microscopic dynamics, %``full network" of reciprocally coupled elements, 
only few representative (stochastic) variables are  
considered by integrating out all other degrees of freedom (d.o.f.). This is the key idea of the celebrated Mori-Zwanzig approach~\cite{Zwanzig1973} yielding a generalized Langevin equation, which involves noise and a memory kernel satisfying a fluctuation-dissipation relation (FDR), and may stochastically describe the motion of a colloid in a complex environment (e.g., a viscoelastic fluid~\cite{Zwanzig2001,Rouse1953,Maes2013,Franosch2011}).

However, the idea of reciprocal couplings and its thermodynamic implications breaks down in many living and artificial complex systems, 
where %
more general interactions, in particular, \textit{non-reciprocal} couplings between mesoscopic subsystems, or (stochastic) d.o.f., naturally emerge~\cite{agudo2019active,Durve2018,Kompaneets2008,Ivlev2015,Saha2020}; 
as, e.g., in pedestrian dynamics~\cite{Helbing1995,Moussaid2011,Karamouzas2014}, in complex plasmas~\cite{Chaudhuri2011,Morfill2009,Lisina2013,Vaulina2015,Bartnick2016b}, or in bio-chemical systems~\cite{Kronzucker2008,bo2015thermodynamic,hartich2016sensory}. Moreover, state-of-the-art experimental techniques %based on videomicroscopy and optical feedback 
enable the realization of almost arbitrary interactions between colloidal particles~\cite{Khadka2018,Geiss2019}, including non-reciprocal ones~\cite{Lavergne2019}. Tuning the interactions opens up the possibility to experimentally explore fundamental principles, and to manufacture artificial systems on the fluctuating scale, like Brownian molecules~\cite{Khadka2018,Geiss2019}. % or colloidal machines \cite{Martinez2016,Blickle2012}. 
Recently, also in quantum systems it was demonstrated that the implementation of non-reciprocal couplings can be used to build new types of devices, e.g., directional amplifiers~\cite{Fang2017,Metelmann2015,Manaselyan2019,Shen2018,Malz2018}.
Further, non-reciprocal couplings between (effective) variables are present in various models for active matter. For example, to describe active self-propelled motion~\cite{ramaswamy2010mechanics,ramaswamy2017active,Fodor2016,marchetti2013hydrodynamics,speck2019thermodynamic}, 
the temporal evolution of the particle's position 
is assumed to be affected by the orientation (due to the flagella or asymmetric flow field), %
but there is no backcoupling.

While some models which involve non-reciprocal interactions have already been studied from a thermodynamic perspective~\cite{Shankar2018,pietzonka2017entropy,Caprini2019,argun2016non,marconi2017heat,Dabelow2019,micali2016bacterial,Hinrichsen2013}, the general thermodynamic and information-theoretical implications of non-reciprocity itself have, to our knowledge, not been discussed so far. 
This is the first major goal of this paper. To this end, we will review and reinterpret some results from the literature (for systems with two d.o.f.), and derive new formulae for larger systems. 
In particular, we consider (mostly) overdamped Markovian systems of $n+1$ non-reciprocally coupled subsystems $X_{0,1,\dots, n}$ %($j=0,1,\dots, n$)
with white noise. Each subsystem can represent, e.g., the position of a colloid in an experiment (accordingly, we will assume that the variables are even under time-reversal, like positions or angles).
By considering different thermodynamic quantities, we investigate the following questions: Can non-reciprocal systems reach a state of thermal equilibrium? Is there a crucial difference between nonequilbrium states induced by non-reciprocity, vs. external drivings? 
Indeed, we show here that, except for some specific cases, non-reciprocal systems are inherently out of equilibrium, even in the absence of external forces or (temperature) gradients.
In order to discuss the fundamental consequences of non-reciprocity on a purely analytical basis, we will consider linear models. However, as we will discuss, several conclusions take over to non-linear models. 
As different representatives of non-reciprocal coupled systems that share some crucial features we will consider, on the one hand, active systems and, on the other hand, feedback-controlled systems.
The second main goal of this paper is to explain why, under certain conditions, a setup with non-reciprocal linear couplings can be used to build a ``microswimmer'', a 
``feedback controller'', or a ``Maxwell demon''.
For microswimmers, thermodynamic notions are already a huge topic~\cite{Shankar2018,pietzonka2017entropy,Caprini2019,argun2016non,marconi2017heat,Dabelow2019,Fodor2016,Fodor2018,micali2016bacterial,Li2019}. Here, we calculate the information and energy flow between the particle (here $X_0$) and its propulsion mechanism (here represented by at least one subsystem $X_1$), confirming general expectations, e.g., the active swimmer heats up its environment but never cools it down.
In contrast, in the context of time- and space-continuous feedback~\cite{Loos2019,Munakata2014,Rosinberg2015, Rosinberg2017,van2019uncertainty}, the connection to non-reciprocal coupling is rather uncommon and new. Therefore, we dedicate a more detailed analysis to this point. We show 
that linear non-reciprocal couplings can be used to construct a time-delayed feedback loop, and clarify under which conditions a non-reciprocal coupled d.o.f. can extract energy from a single heat bath, making it a ``Maxwell demon". We further find conditions under which thermal fluctuation suppression (or enhancement), i.e., ``isothermal compression or expansion'' of a single-particle gas are possible. 

While some of the questions and connections discussed here may seem to be intuitively clear, almost representing ``common wisdom'', 
there are only few studies where these issues are formally addressed. Moreover, we also detect counter-intuitive phenomena. 
For example, non-Markovian processes can exhibit a nonequilibrium steady state (NESS) without dissipation, where the entropy is exported \textit{purely} in the form of information, implying that information and entropy are transported without accompanying energy flow (while in total sustaining this process relies on external energy supply).
Furthermore, we show that, under certain conditions, a system of two isothermal subsystems with non-reciprocal coupling can be mapped onto a reciprocal system with a temperature gradient, building a bridge to other active matter models~\cite{Li2019,roldan2018arrow,netz2018fluctuation}. In this context, we also consider the underdamped case. In addition, we provide a detailed derivation of the relevant information flows, which is, so far, a quantity that is not well-established for time- and space-continuous systems. 

From a conceptual viewpoint it is important to also think about situations, where a portion of the d.o.f. might not be invisible to a (``marginal'') observer. Even more, in some theoretical models, % (e.g., for a microswimmer within the AOUP model), 
a portion of the d.o.f. has no direct physical interpretation. Then, the dynamics can be equivalently formulated as a non-Markovian, one-variable equation (for $X_0$) with a memory kernel and colored noise, upon projecting out $X_{j>0}$.
In such a situation, the interpretation of thermodynamic quantities must be treated with care, and is indeed subject of a recent debate~\cite{Shankar2018,Dabelow2019,Caprini2019,Rosinberg2015,Rosinberg2017}. 
To account for this fact, we will pay special attention to the different measures of (non)equilibrium on the levels of the Markovian and non-Markovian description, and also explicitly consider the entropy balance of an individual subsystem. We will further comment on the connection to so-called
%An interesting line of research is the search for 
``effective thermodynamic'' descriptions~\cite{Herpich2020,Polettini2017}.

We close this introduction with a brief outline. After introducing the model in Sec.~II, we will investigate under which conditions detailed balance and the fluctuation-dissipation relation are satisfied (Sec.~III). Then, we will calculate the total entropy production of the entire system and the dissipation of an individual subsystem in Sec.~IV. Thereafter we will consider the entropy balance of an individual sub-system and derive explicit expressions for the information flows through the system (Sec.~V). In Sec.~VI, we show that, under certain conditions, a non-reciprocal (overdamped) system can be mapped onto a reciprocal one. This is also possible for the corresponding underdamped case, as discussed in Sec. \ref{sec:underdampedCases}. There, we also consider the heat flow in a non-reciprocal system with inertia. We finally conclude in Sec.~VII.

%%%%%%%%%%%%%%%%%%%%%%%%%%%%%%%%%%%%%%%%%%%%

%

%%%%%%%%%%%%%%%%%%%%%%%%%%%%%%%%%%%%%%%%%%
%
%
%      
%                     Main part
%
%
%%%%%%%%%%%%%%%%%%%%%%%%%%%%%%%%%%%%%%%%%%%%
\section{Model}\label{SEC:Model}
\begin{figure}
\begin{minipage}[l]{0.3\textwidth}
\begin{flushleft} (a)\end{flushleft}
\end{minipage}
\begin{minipage}[l]{0.3\textwidth}
\begin{flushleft} (b)\end{flushleft}
\end{minipage}
\begin{minipage}[t]{0.3\textwidth}
\begin{flushleft} (c)\end{flushleft}
\end{minipage}
\\[-0.8cm]
\begin{minipage}[t]{0.3\textwidth}
%\textcolor{white}{.}\vspace{-0.3cm}\\
	\includegraphics[width=0.6\textwidth]{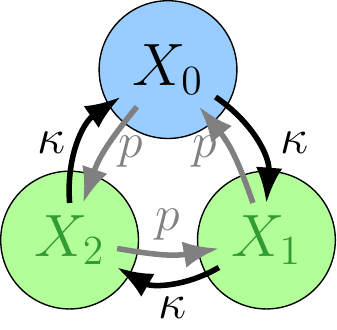}
\end{minipage}
\begin{minipage}[t]{0.3\textwidth}
	\includegraphics[width=0.8\textwidth]{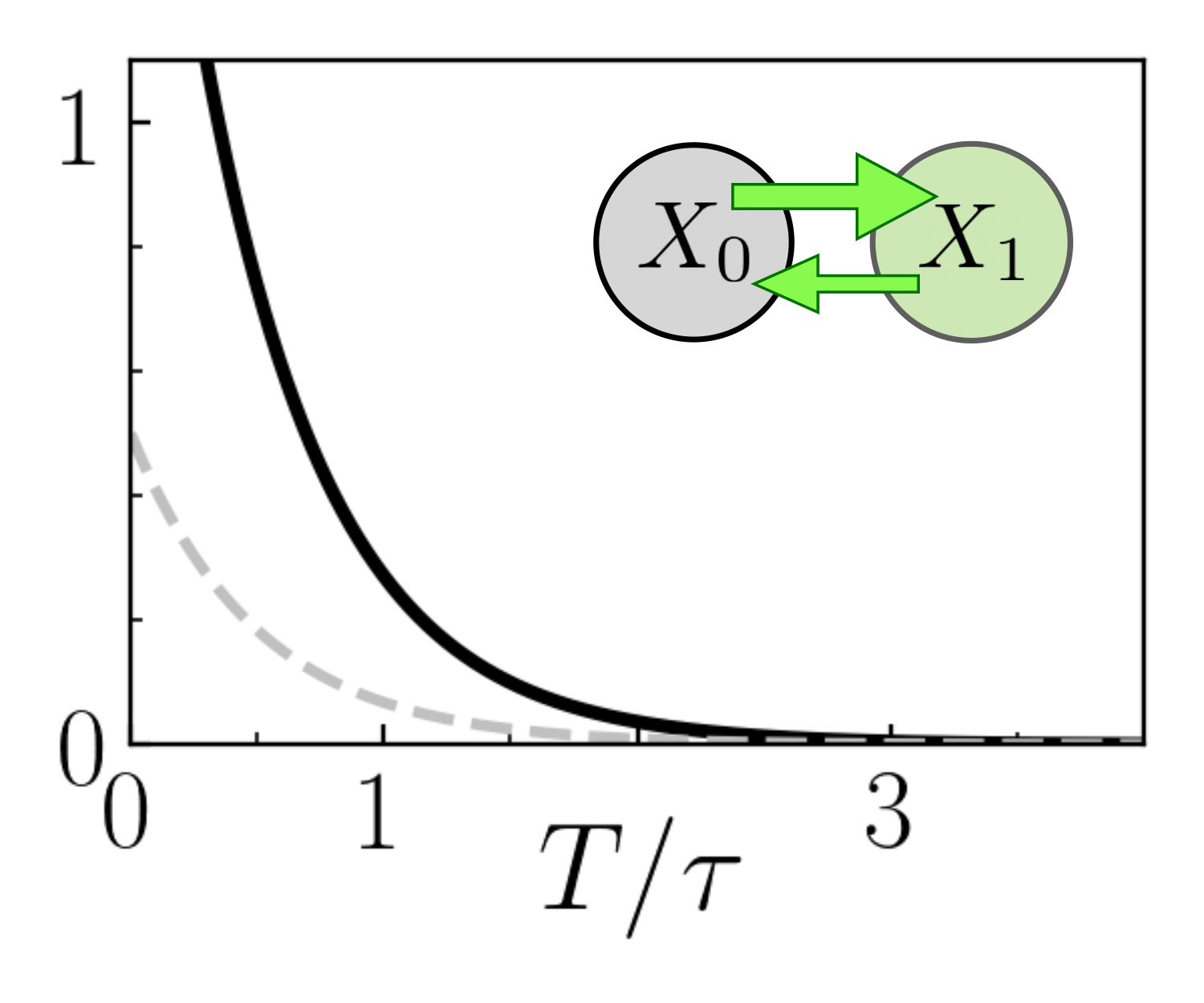}
\end{minipage}
\begin{minipage}[t]{0.3\textwidth}
	\includegraphics[width=0.8\textwidth]{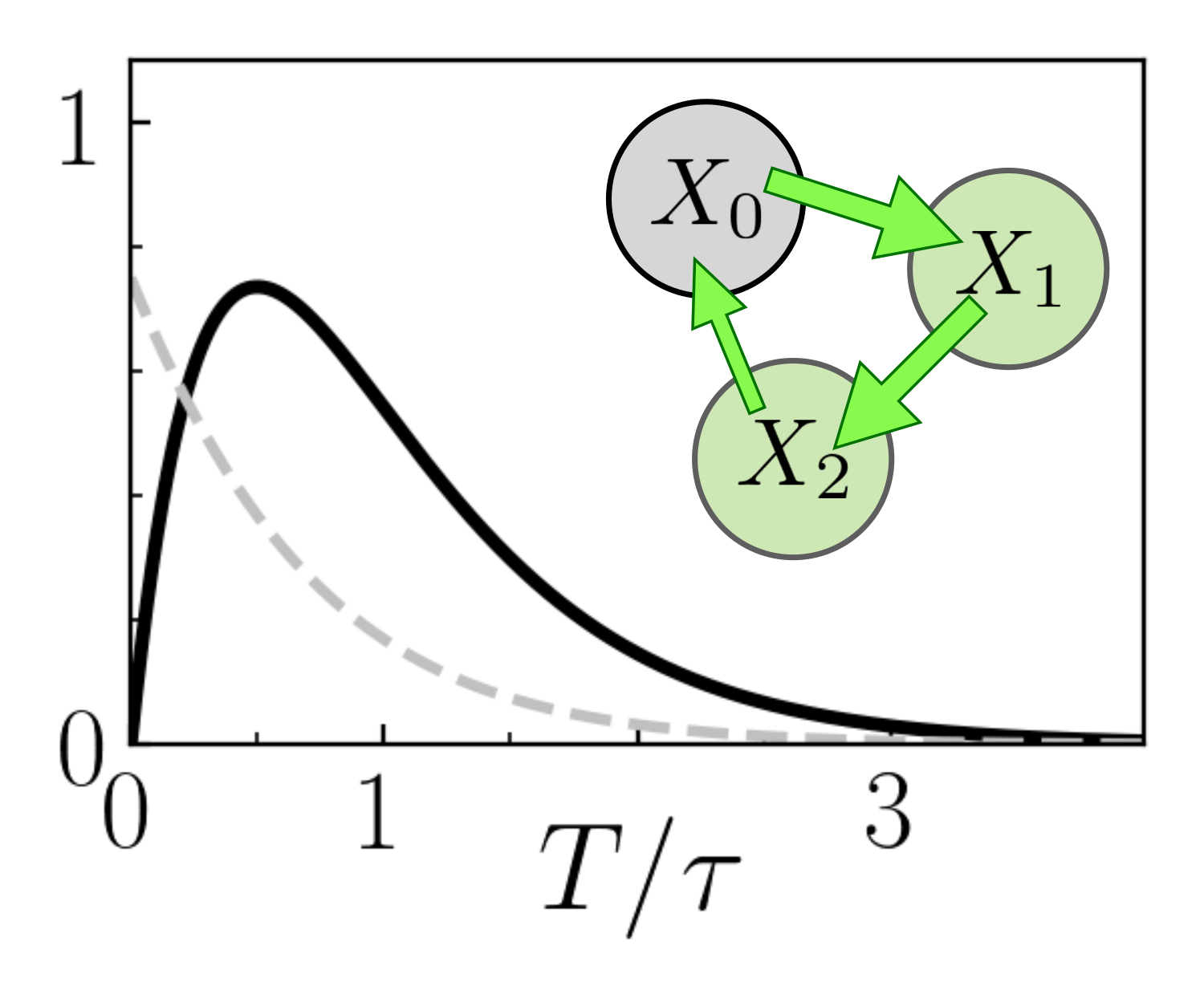}%{Figures/Ring_n=2.pdf} %
\end{minipage}
  	\caption{(a) Ring of three d.o.f.. %, defined by~(\ref{eq:Network}) with $n=2$, and $a_{jj} =p + \kappa$, $a_{j j+1}= -p$, $a_{j+1 j} = -\kappa$, $\gamma_j=1$. 
     		For reciprocal coupling $\kappa=p$, when this corresponds to a mechanical system, the memory kernel is exponentially decaying. Non-reciprocal coupling $\kappa\neq p$ yields non-monotonic memory~(\ref{eg:kernel_n=3}).
             %\label{fig:nonmonotonic}
(b,\,c) Memory kernels $K(T)$ (solid black lines) and noise correlations $\nu_n$ (grey dashed lines) generated by systems with the coupling topologies as shown in the insets, (b) $n=1$, (c) $n=2$, and $k=1$, $\tau=1/2$, $\mathcal{T}_{j>0}=1$. 
	} \label{fig:kernels}
  
\end{figure}
%

%\subsection{Markovian formulation}
We consider time- and space-continuous systems described by Markovian overdamped Langevin equations (LE)
%\begin{subequations}\label{eq:Network}
\begin{align}\label{eq:Network}
 \underline{\underline{\gamma}}  \underline{\dot{X}}
   = 
   %\underline{\underline{a}}
     \begin{pmatrix}
a_{00}  &  a_{01} &  ... &  a_{0n} \\
a_{10}  &  a_{11} &  ... &  a_{1n} \\
& &\vdots &\\
a_{n0}  &  a_{n1} &  ... &  a_{nn}\\
\end{pmatrix}
\underline{X}
   +
     \begin{pmatrix}
     f_0 \\
     0\\
     \vdots \\
     0\\
      \end{pmatrix}
      +
     \,
       \begin{pmatrix}
       \xi_0 \\
       \xi_1 \\
       \vdots \\
       \xi_n \\
        \end{pmatrix}
\end{align}
%\end{subequations}
with the vector
$\underline{X}=(X_0,X_1,...,X_n)^T \in \mathbb{R}^{n+1}$ involving $n+1$ stochastic d.o.f..
We will discuss thermodynamic properties of both, the entire system $\{X_0,X_1,...,X_n\}$, and of the individual $X_j$. 
To set the focus, we will occasionally call
$\{X_0,X_1,...,X_n\}$ the ``super-system'', while an individual $X_j$ will be called a ``sub-system''.	
Further, $\xi_j$ denote zero-mean, Gaussian white noises with $\langle \xi_i(t)\xi_j(t') \rangle =2 k_\mathrm{B}\mathcal{T}_j \gamma_j \,\delta_{ij}\delta(t-t')$ at temperatures $\mathcal{T}_j\geq 0$,
$ j\in \{0,1,...,n\}$, with $k_\mathrm{B}$, $\gamma_j$ being the Boltzmann and friction constants that also appear in the diagonal friction matrix $\underline{\underline{\gamma}}$ with $\gamma_{jj}=\gamma_j$. $f_0$ is an, in general, nonlinear force.
The coupling matrix $\underline{\underline{a}}$ defines the strength of the couplings $a_{ij}$, and gives the timescale $\gamma_j/a_{jj}$ of the {exponential relaxation dynamics} of each d.o.f., due to the restoring forces $a_{jj}X_j$. We will focus on cases where the motion of $X_0$ is \textit{confined}, i.e., $a_{00}<0$, and consider
	natural boundary conditions, i.e., the probability to find the particle vanishes at $X_0 \to \pm \infty$. 
Further, we will focus on situations, where a stable steady state exists, which is the case whenever the real part of the largest eigenvalue of the coupling matrix $\underline{\underline{a}}$ is negative.

At this point, we may already note one apparent difference between reciprocal system ($a_{ij}= a_{ji}~\forall i,j$) and those that involve \textit{non-reciprocal couplings} ($a_{ij}\neq a_{ji}$), that is, 
only the purely reciprocal coupled equations can be expressed as derivatives of a Hamiltonian, plus noise terms (and, if present, plus non-conservative forces $f_0$). In that case, (\ref{eq:Network}) can be written as $\gamma_j \dot{X}_j = -\frac{\partial \mathcal{H}}{\partial X_j} +\xi_j$, with the Hamiltonian
\begin{align}\label{Hamiltonian} %
\mathcal{H}= \sum_{j=0}^{n}\big[ V_j(X_j) +  \sum_{i > j} H_\mathrm{int}(|X_i-X_j|) \big] =\sum_{j=0}^{n} \big[\frac{- a_{jj}}{2}X_i^2 + \frac{{-} a_{ij}}{2} \sum_{i> j} (X_i-X_j)^2 \big],
\end{align}
%\red{ Such systems have been studied in .... }
where the last term in (\ref{Hamiltonian}) represents the interaction part, $H_\mathrm{int}$.
In contrast, non-reciprocal couplings appear as a non-conservative force (like $f_0$). In that case, (\ref{eq:Network}) corresponds to $\gamma_j \dot{X}_j = -\frac{\partial V_j}{\partial X_j} + \sum_{i\neq j} a_{ij} X_i  +\xi_j$.
%, which only exists for reciprocal couplings.

Equivalently to (\ref{eq:Network}), one can describe the dynamics of one d.o.f., say $X_0$, by a one-variable LE
\begin{align} \label{eq:LE-X0}
\gamma_0\dot{X}_0(t) = \,&a_{00}X_0(t) + \int_{0}^{t} K(t-t') X_0(t')\mathrm{d}t'
%\nn 
%&
+ f_0+ \nu(t)  + \xi_0(t),
\end{align} 
which can be derived by projecting the $X_{j>0}$ onto $X_0$, 
as described in~\cite{Loos2019b,Zwanzig1973} and in Appendices \ref{sec:MemoryRing} and \ref{sec:deriveKandNu}.
Generally (unless the time-scales of $X_0$ and $X_{j>0}$ are well-separated), (\ref{eq:LE-X0}) is a 
\textit{non-Markovian} LE, i.e., it comprises memory. In particular, it involves a time-nonlocal force
depending on the \textit{past} trajectory, weighted with a memory kernel $K$, and $\nu$ is a zero-mean, Gaussian colored noise (both depend on the topology of the coupling matrix, concrete examples are given below). For $\mathcal{T}_{j>0}\equiv 0$, there is no colored noise in~(\ref{eq:LE-X0}). We aim to emphasize that the dynamics of $X_0$ is \textit{identical} to~(\ref{eq:Network}). Using (\ref{eq:LE-X0}) instead of (\ref{eq:Network}) can be regarded as a coarse-graining or marginalization, because the dynamics of $X_j$ is not explicitly considered. However, it does \textit{not} imply loss of information about, or approximation of, $X_0$. {One should note that, in reverse, for a non-Markovian process (\ref{eq:LE-X0}), a corresponding Markovian representation~(\ref{eq:Network}) is not unique. Thus, a specific memory can be realized by different Markovian networks [this can be seen, e.g., from Eq.~(\ref{eq:ExampleI}) by the fact that $a_{01}$ and $a_{10}$ only arise as product, $a_{01}a_{10}$].}
%\subsection{Non-Markovian formulation}

For the sake of generality,
we deliberately do not focus on a specific model, and rather offer different interpretations for the involved d.o.f.; explicit examples will be given below. 
However, a situation of special interest is that the observer only sees parts of the system (say only $X_0$), while the other d.o.f. are ``hidden''. Even more, in some cases, only certain d.o.f. (say only $X_0$), represent actual, physical d.o.f. (such as the position of a colloid), whereas the others (say $X_{j>0}$) {are effective (or auxiliary) variables representing} those parts of the complex environment which generate a feedback loop or active motion.
In such a situation, a non-Markovian description~(\ref{eq:LE-X0}), which only involves $X_0$, may be the more fundamental one. We will discuss both situations, only $X_0$ or all $X_j$ being observed, in this paper.

Before we start with investigating the thermodynamic consequences of non-reciprocity, we first aim to discuss the relationship between non-reciprocal coupling in (\ref{eq:Network}) and resulting memory in~(\ref{eq:LE-X0}) and then give some examples for systems that can be modeled by (\ref{eq:Network}) and (\ref{eq:LE-X0}). 
%
%%%%%%%%%%%%%%%%%%%
%
%
% % % % % % % % % % % % % % % % % %
%. 
%
\subsection{Memory induced by non-reciprocally coupled systems}
We begin by considering the smallest version of (\ref{eq:Network}) with $n=1$. 
{While various aspects of this case have been studied previously~\cite{Shankar2018,Dabelow2019,Caprini2019,Mandal2017,Bonilla2019,Crisanti2012,Puglisi2009}, the full implications of non-reciprocity have so far, to the best of our knowledge, not been discussed.} 
For $n=1$, the memory kernel $K$ and the noise correlations $C_\nu(T) := \langle \nu(t)  \nu(t+T) \rangle$ are both found to decay {exponentially} for reciprocal as well as non-reciprocal coupling, and read
\begin{align} 
%\begin{cases}
K(T) &=   (a_{01}a_{10} /\gamma_{1}) e^{a_{11}T/\gamma_1},\nn
C_\nu(T)  &=  k_\mathrm{B}\mathcal{T}_1 (a_{01}^2/a_{11}) e^{a_{11}T/\gamma_1}.
\label{eq:ExampleI} %\end{cases}
%\label{eq:ExampleI}%\tag{I}
\end{align} 
An exemplary plot of both functions is given in Fig.~\ref{fig:kernels}\,(a).

Let us now investigate the effect of adding more sub-systems $X_j$ to the super-system~(\ref{eq:Network}), such that there may be an interplay of multiple non-reciprocal interactions.
Most importantly in the present context, this leads to complex types of memory beyond the single exponential decay.
To illustrate this, let us consider 
%the case $n=2$
a ring of three d.o.f., %
where all (counter-)clockwise couplings are set to ($p$) $\kappa$, (with $a_{j j-1} =\kappa$, $a_{j j+1}= p$, $-a_{jj} =p+\kappa$), %while counterclockwise are set to
as sketched in Fig~\ref{fig:kernels} (a). This super-system % version 1
%, which
generates the memory kernel
\begin{align}
K(t-t')=
%...\kappa
\frac{e^{\left(-\sqrt{p \kappa}+p+\kappa\right) t'-t \left(\sqrt{p \kappa}+p+\kappa\right)}}{2 \sqrt{p \kappa}}
%\times
%\nn
\left[ \left(p^{3/2}+\kappa^{3/2}\right)^2 e^{2 \sqrt{p \kappa} t}
-\left(p^{3/2}-\kappa^{3/2}\right)^2 e^{2 \sqrt{p \kappa} t'}\right], \label{eg:kernel_n=3}
\end{align}
(see Appendix~\ref{sec:MemoryRing} for a derivation). 
%In the same manner, w
%
For {reciprocal}, i.e., conservative couplings, $\kappa=p$, 
%$a_{ij}=a_{ji}$, 
%where the super-system corresponds to a mechanical model,
(\ref{eg:kernel_n=3}) simplifies to an exponential decay
%. For example, if all couplings are identical, 
$K(T=|t-t'|) =2\kappa^2 e^{-\kappa T} $. 
In contrast, if the coupling is non-reciprocal, we find that the super-system~(\ref{eq:Network}) generates a \textit{non-monotonic} memory kernel, despite the linearity of all couplings. 
In the present example, the memory kernel (\ref{eg:kernel_n=3}) has a maximum at a finite time difference. In the limit of \textit{unidirectional} coupling $p\to 0$, 
the memory kernel~(\ref{eg:kernel_n=3}) converges to a Gamma-distribution
$K(T)= \kappa^3\, T e^{- \kappa T },$ which has a pronounced maximum near $\kappa/3$, see Fig.~\ref{fig:kernels}\,(b). Noteworthy, in this limit, the kernel vanishes at $T=0$, i.e., the instantaneous position does not contribute to the integral $\int X_0(t')K(t-t')\mathrm{d}t'$ in (\ref{eq:LE-X0}) [while the integral is dominated by the instantaneous position for reciprocal coupling].
In Appendix~\ref{sec:MemoryRing}, we discuss the general case where all couplings are different,
yielding very cumbersome expressions
while the overall characteristics are the same.

Playing around with different coupling topologies and system sizes, we generally find that 
non-reciprocal coupling is a crucial ingredient to generate non-monotonic memory, while reciprocal couplings always yield monotonic kernels. With an appropriate coupling topology, it is also possible to generate memory kernels with multiple maxima. We observe that a kernel with $n$ extrema can be represented via (at least) $n$ d.o.f..

On the other hand, we observe that the correlation $C_\nu(T)$ of the colored noise produced by Markovian systems with ring topology [of type (\ref{eg:kernel_n=3})] is always monotonically decreasing with $T$ (see, e.g., Fig.~\ref{fig:kernels}). This implies a broken fluctuation-dissipation relation, as we will discuss below in Sec~\ref{sec:FDR}. For other coupling topologies, linearly and non-reciprocally coupled d.o.f. can also induce non-monotonic noise correlations. A systematic study of the connections between coupling topology, generated memory, and the resulting correlation functions will be presented in~\cite{Doerries2020}.%

% % % % % % % % % % % % % % % % % % % % % % % % % % % % % % % % % % % % % % %
\subsection{Examples}
\begin{figure}
	\centering
	\includegraphics[width=0.51\textwidth]{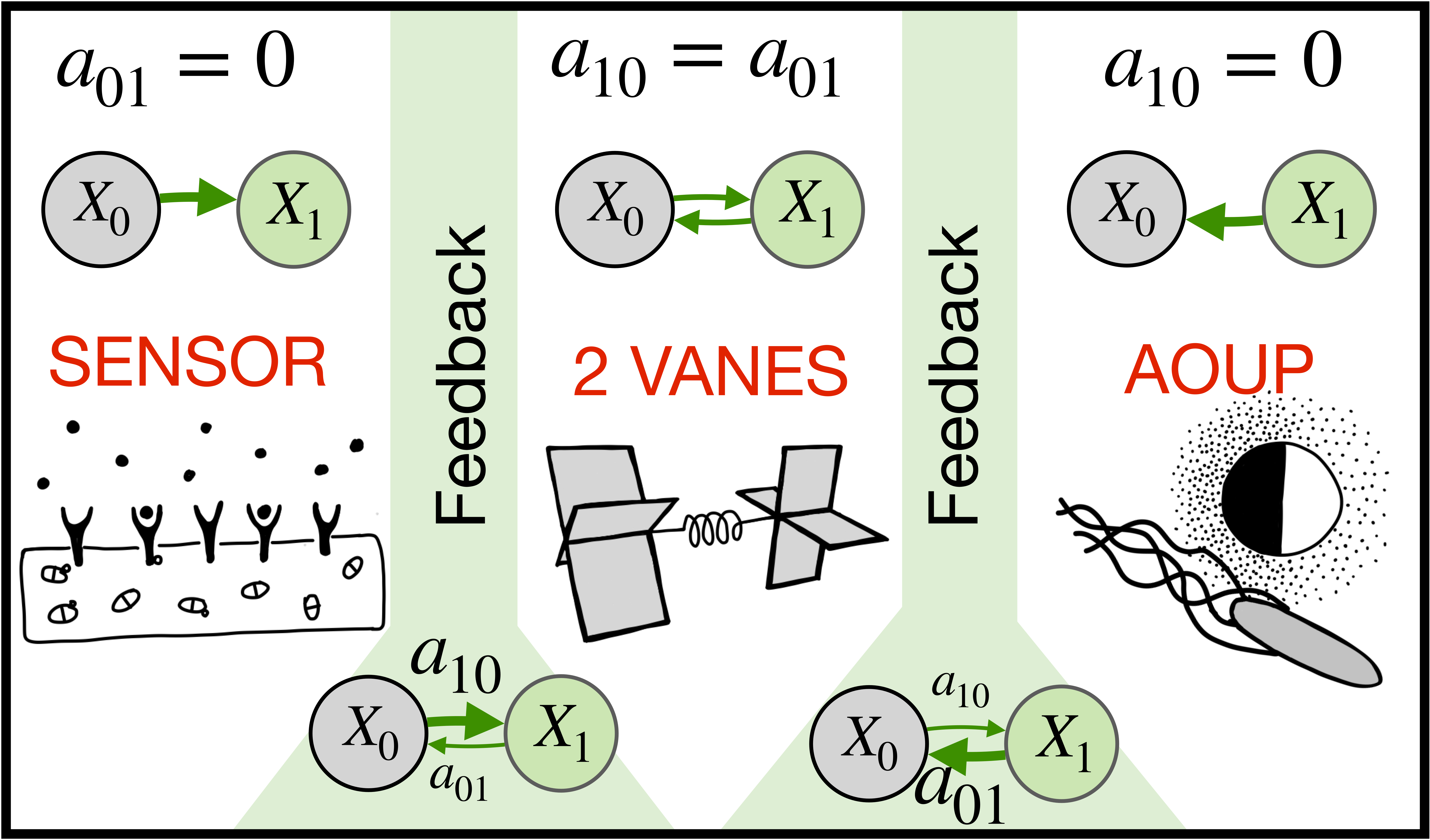} %{OverviewArrow-crop.pdf} %{Figures/Kernels_insets-crop.pdf} %
	\caption{
		Overview of various systems describable by the generic model~(\ref{eq:Network}) with $n=1$.
		%, with increasing ratio $a_{01}/a_{10}$ from left to right. 
		\textit{Left:} 
		%If $X_1$ is unidirectionally coupled to the system d.o.f. $X_0$, i.e., $a_{01}=0$, such that only $X_1$ ``sees'' $X_0$, but not \textit{vice versa}, 
		for unidirectional coupling ($a_{01}=0$), 
		$X_1$ corresponds to a ``cellular sensor'' in the model~\cite{hartich2016sensory}, and $X_0$ to a d.o.f. measured by that sensor (e.g., a ligand concentration), see Eq.~(\ref{eq:Sensor}) and text below. \textit{Center:} for reciprocal coupling ($a_{01}=a_{10}$), $X_{0,1}$ correspond to the angles of two mechanically coupled vanes~\cite{Sekimoto2010}. \textit{Right:} For unidirectional coupling ($a_{10}=0$), $X_0$ corresponds to the position of a microswimmer within the AOUP model, while $X_1$ represents the self-propulsion velocity~\cite{Shankar2018,Dabelow2019,Caprini2019,Mandal2017,Bonilla2019,Martin2020}, see (\ref{eq:AOUP}) and below. In the intermediate cases with bidirectional non-reciprocal coupling, $X_1$ corresponds to a feedback controller acting on a colloid at position $X_0$.
	} \label{fig:overview}
\end{figure}
Let us now consider some exemplary systems of type~(\ref{eq:Network}) with non-reciprocal interactions. We start with a brief summary of models known from the literature and then introduce our new models with feedback. Figure~\ref{fig:overview} provides an overview for the case $n=1$. 
 
For {reciprocal} coupling, the dynamics of the two d.o.f. $X_{0}$ and $X_{1}$, corresponds to the {angles of two vanes} that rotate in two different heat baths at $\mathcal{T}_0$ and $\mathcal{T}_1$, and are coupled by a torsion spring with spring constant $a_{01}=a_{10}$.
At $\mathcal{T}_0 \neq \mathcal{T}_1$ this setup was considered as a minimal model for {heat conducting} through mechanical motion, as discussed in~\cite{Sekimoto2010} (see p.\,154). 
Further, for {unidirectional} coupling $a_{10}=0$, $a_{01}> 0$, $a_{11}<0$, 
 our model (\ref{eq:Network}) reduces to the active Ornstein-Uhlenbeck particle (AOUP) model with transitional noise~\cite{Shankar2018,Dabelow2019,Caprini2019,Mandal2017,Bonilla2019,Martin2020}, reading
 \begin{align}\label{eq:AOUP}
 	\gamma_0 \dot{X}_0 = a_{00}X_0 + f_0 + X_1 +\xi_0,~~~\dot{X}_1 =  -X_1/\tau +\xi_1/(\gamma_1 \tau),
 	 \end{align}
 which corresponds to (\ref{eq:Network}) with $\tau=\gamma_1/|a_{11}|\geq 0$, and $a_{01}=1$.
 This is a simple (overdamped) model for active swimmers, where $X_0$ corresponds to the position of a \textit{microswimmer} in an harmonic trap with stiffness $a_{00} \leq 0$, while $X_1$ represents the ``self-propulsion velocity"~\cite{Martin2020}, pushing $X_0$ away from it. In a real system, the propulsion could be created by the flagella of a bacterium, or the asymmetric flow field around a Janus colloid. In the corresponding non-Markovian representation (\ref{eq:LE-X0}), the colored noise [which is here the only type of memory, as $K\equiv 0$ when $a_{10}=0$] yields the \textit{persistence} of the motion, and $\tau$ quantifies the ``persistence'' of the ``active noise''~\cite{Martin2020}. 
 Next, the super-system with reversed unidirectional coupling (i.e., $a_{01} \leftrightarrows a_{10}$), was recently suggested as a model for a {cellular sensor}~\cite{hartich2016sensory}
   \begin{align}\label{eq:Sensor}
\gamma_0\dot{X}_0 = a_{00}  X_0 +\xi_0 ,~~~\dot{X}_1 = (a_{11}/\gamma_1) [X_1 -X_0 ]+\xi_1/\gamma_1, 
   \end{align}
   with $ a_{00} < 0$, $a_{11} < 0$, which corresponds to (\ref{eq:Network}) with $a_{11} = -a_{01} < 0$. 
   Thereby, the cellular sensor is described by a one-dimensional variable $X_1$ (giving the state of the sensor at time $t$, which is, according to~\cite{hartich2016sensory}, related to the number of bound receptors). The purpose of the sensor is to measure a certain external d.o.f., $X_0$, which could be the concentration of some ligand~\cite{hartich2016sensory}.
 %bo2015
 Last, we aim to note that the model for a {cellular sensor} with memory from Ref.~\cite{hartich2016sensory}, 
	corresponds to the case $n=2$, where $X_2$ represents the past state of the sensor, i.e., the memory (related to the number of phosphorylated internal proteins~\cite{hartich2016sensory}). Then, $X_0 \to X_1$, and $X_1 \to X_2$ are coupled unidirectionally, and there is no direct link between $X_0$ and $X_2$. 

As we will show in this paper, the generic system~(\ref{eq:Network}) with non-reciprocal couplings also includes cases where the d.o.f., $X_{j>0}$, can be regarded as a \textit{feedback controller} continuously operating with the force $F_\mathrm{c}$ on a system $X_0$, yielding a dynamical equation of the colloid
\begin{align} \label{eq:LE-X0_Feedback}
\gamma_0\dot{X}_0 = \,&a_{00}X_0 + F_\mathrm{c}
%%\nn 
%%&
+ f_0+ \nu  + \xi_0,
\end{align} 
which is a special type of~(\ref{eq:LE-X0}).
A characteristic aspect of feedback control is the occurrence of a \textit{time delay} between ``measurement'' and ``control action''. In experimental setups, this delay either emerges naturally due to finite signal transmission or information processing times (e.g., think of optical feedback with the help of videomicroscopy~\cite{Bechhoefer2005,Loos2019,debiossac2019thermodynamics,Wallin2008,Balijepalli2012}), or may be implemented intentionally (e.g., in Pyragas control~\cite{Schoell2008,Pyragas1992}), because it is known to induce interesting dynamical and thermodynamical behavior, such as particle oscillations~\cite{Bechhoefer2005,Schneider2013,Loos2014}, transport~\cite{Loos2014}, or a net energy extraction from the bath~\cite{Loos2019}.
The controller model with $n=1$ and bidirectional non-reciprocal coupling can be interpreted as a minimal realization of such a controller. However, it yields an exponentially distributed delay with maximum at $t-t'=0$. In contrast, the feedback loop often has a typical finite duration, i.e., the control action depends on $X_0(t-\tau)$, with a distinct characteristic delay time, $\tau>0$, implying that the equation of the controlled system (here $X_0$) involves a memory kernel with a maximum around $\tau$. It now becomes clear that a unidirectional ring with %$n\geq 1$. 
$n=2$ can describe such a controller with preferred delay time. Specifically, setting
$a_{1 0}=a_{2 1}=-a_{11}=-a_{22}=\gamma_1 /\tau$, $\mathcal{T}_{1}=\mathcal{T}_2$, $\gamma_{1}=\gamma_2$, $k=a_{0n}$ 
yields a kernel
\begin{align}% Original result (consistent with PRX) K(T) &= a_{0n}\frac{4 \, T }{\tau^2}  e^{-\frac{2T}{\tau} }.
K(T) &= (k /{\tau^2}) T \,e^{- {T}/{\tau} },\nn
C_\nu(T)    &=  [k_\mathrm{B}\mathcal{T}_1/(2\gamma_1)]  k^2  
\left( 
3 \tau +
%p=1 l=1
T  \right) \,e^{- T /\tau},
\label{K_noise_II}
%\end{aligned}
%\tag{II}
\end{align}
with a pronounced maximum at $\tau $. The feedback force in the non-Markovian equations (\ref{eq:LE-X0_Feedback}) or (\ref{eq:LE-X0}) is $F_\mathrm{c}=\int_{0}^{t} X_0(t')K(t-t')\mathrm{d}t'$, and, in the Markovian description the feedback force is $k X_n$, respectively. Note that, due to this setting, the only remaining free controller parameters are the time delay $\tau$ and the feedback gain $k$. 
To better compare the controllers with $n=1$ and $n=2$, we analogously set $a_{10}=-a_{11}=\gamma_1/\tau$, and $a_{01}=k$ in the case with $n=1$, obtaining from~(\ref{eq:ExampleI}),
\begin{align} 
%\begin{aligned}
K(T) &=   (k /\tau) \,e^{-T/\tau},\nn
C_\nu(T)  &=  (k_\mathrm{B}\mathcal{T}_1/\gamma_1) k^2\tau\, e^{-T/\tau}.
\label{eq:Controller_n=1}
\end{align} 
In this paper, we focus on the cases $n=1,2$, a generalization towards higher $n$ will be discussed in~\cite{Loos2020c}. We note that the limit $n\to \infty$ yields a $\delta$-distributed memory kernel around $\tau$~\cite{Loos2019,Loos2019b}, i.e., $K\propto \delta(T-\tau)$. Such stochastic delay differential equations are infinite-dimensional, which makes their treatment very involved, especially when it comes to thermodynamics~\cite{Rosinberg2015,Rosinberg2017,Rosinberg2018,Loos2020}. In comparison, the model proposed here has in total three d.o.f. and is thus, quite handy.

% % % % % % % % % % % % % % % % % % % % % % % % % % % % % % % % % % 
\section{Intrinsic non-equilibrium}\label{sec:noneq}
Now we turn to the thermodynamic properties induced by the occurrence of non-reciprocal interactions, focusing on the long-time behavior $t\to\infty$, when transient dynamics due to the initial conditions have decayed and the system has approached a {steady state}.%

% % % % % % % % % % % % % % % % % % % % % 

We start by clarifying whether thermal equilibrium can exist despite non-reciprocity. As mentioned before, non-reciprocal interactions are non-conservative. One might therefore guess that a system with non-reciprocal interactions cannot reach thermal equilibrium. To investigate this question, we check the detailed balance (DB) condition on the level of the Markovian representation~(\ref{eq:Network}). Since the latter is only meaningful when all d.o.f. have a physical interpretation, we also discuss the fluctuation-dissipation relation (FDR) on the level of the non-Markovian description~(\ref{eq:LE-X0}).

Since we are interested in analytical solutions, we will focus on the linear case, i.e., $f_0=0$. We stress, however, that the framework is readily adaptable to cases where a nonlinear force act on $X_0$, then requiring numerical solutions.

\subsection{Detailed Balance}\label{SEC:DB}
To investigate whether the super-system~(\ref{eq:Network}) can approach thermal equilibrium, we check the detailed balance condition. 
%i.e., whether the $(n+1)$-dimensional probability current vanishes.To access this quantity, 
To this end, we consider the 
flow of the 
$(n+1)$-point joint probability density function (pdf), $\rho_{n + 1}( \underline{x},t )$, of $\underline{x}=(x_0,...,x_n)^T$.
To access this quantity, we utilize the closed, {multivariate} Fokker-Planck equation (FPE)~\cite{Loos2019b} corresponding to (\ref{eq:Network}), which reads
%, whose existence is a key benefit of a Markovian representation. [There is in general no (closed) FPE for non-Markovian systems of type~(\ref{eq:LE-X0}).] 
%
%
\begin{align}\label{eq:FPE-vector}
%\partial_t \rho_{n+1}(\underline{x}) =   -\nabla \underline{J},
\partial_t \rho_{n+1}(\underline{x}) =   -\nabla \underbrace{
 [\underline{\underline{\gamma}}^{-1}\underline{\underline{a}}\underline{x}- \underline{\underline{D}}\nabla ]\rho_{n+1}(\underline{x})
}_{=\underline{J}},
\end{align} 
with the probability current $\underline{J}$ and diagonal diffusion matrix $\underline{\underline{D}}_{jj}= k_\mathrm{B}\mathcal{T}_j/\gamma_j$. We note that $\underline{J}$ is generally constant in steady states, and zero in equilibrium. %\red{It can also be used to quantity the distance from equilibrium \cite{Li2019}.}
Using the identity $ \partial_x \rho= [\partial_x \ln(\rho)] \rho$, we rewrite~(\ref{eq:FPE-vector}) as
$
\partial_t \rho_{n+1} =   -\nabla
	\left[\underline{v} \rho_{n +1}  \right] ,
$
with the $(n+1)$-dimensional phase space velocity~\cite{Weiss2003}
\begin{align}\label{eq:phaseSpace}
\underline{v} =  \underline{\underline{\gamma}}^{-1}\underline{\underline{a}}\underline{x}- \underline{\underline{D}}\nabla \ln \rho_{n+1}(\underline{x}),
\end{align} which is connected to the probability current by $\underline{J}=\underline{v}\rho_{n+1}$. 
DB means that all probability currents vanish, hence, $v_j = 0~\forall j$. From (\ref{eq:phaseSpace}), we obtain the condition $    \underline{\underline{D}}^{-1}\underline{\underline{\gamma}}^{-1}\underline{\underline{a}}\underline{x} = \nabla \ln \rho_{n+1}$, which implies that the vector $\underline{\underline{D}}^{-1}\underline{\underline{\gamma}}^{-1}\underline{\underline{a}}\underline{x}$ is the gradient of a scalar function. This, in turn, is true if and only if $\nabla \times (\underline{\underline{D}}^{-1}\underline{\underline{\gamma}}^{-1}\underline{\underline{a}}\underline{x}) = 0$. Noting that $\underline{\underline{\gamma}}$ and $\underline{\underline{D}}$ are diagonal, this brings us to
\begin{align}\label{eq:Fulfill-DB}
 a_{ij} \mathcal{T}_j =a_{ji} \mathcal{T}_i, 
\end{align}
for all pairwise coupling constants between every two mutually coupled sub-systems. We stress that this condition is irrespective of the coupling topology, or system size.
Remarkably, (\ref{eq:Fulfill-DB}) shows that non-reciprocal systems that fulfill DB do exist, as long as $a_{ij}a_{ij}>0$. However, unidirectional super-systems are by construction pure nonequilibrium models, including the (AOUP) microswimmer, or the controller with non-monotonic memory ($n=2$), see Eq.~(\ref{K_noise_II}).

Condition~(\ref{eq:Fulfill-DB}) further implies that non-reciprocal systems can reach equilibrium despite $\mathcal{T}_i \neq \mathcal{T}_j$. Below, we will show that also the total entropy production vanishes at this point, as well as the heat and information flows [see Eqs.~(\ref{eq:Stot_n=1},~\ref{eq:Q0_n=1},~\ref{eq:Info-flow-0_general})].
This is in sharp contrast to reciprocally coupled (or ``passive'') systems, which generally never equilibrate when being simultaneously coupled to heat baths of different temperatures. This has been shown, e.g., in \cite{Parrondo1996,Hondou2000}. We, however, do not think that our results contradict the previous findings, which exclusively refer to 
	reciprocally coupled systems, like mechanical ones. The non-reciprocal coupling considered in this paper does not correspond to a mechanical coupling, and is typically only realizable with the help of some external apparatus acting on the system (for an example of non-reciprocal coupling realized by light, see~\cite{Lavergne2019}).

\subsection{Fluctuation-dissipation relation}\label{sec:FDR}

Let us now turn to the corresponding non-Markovian process (\ref{eq:LE-X0}) in $x_0$-space, which is more appropriate for models where $X_{j>0}$ have no direct physical interpretations or if a marginal observer only sees $X_0$. On this level of description, the definition of a probability current is less clear, as there is, in general, no corresponding closed FPE~\cite{Loos2019b}. However, from the non-Markovian LE~(\ref{eq:LE-X0}) [at $f_0=0$] alone, we can immediately deduce that the probability current in this {marginalized} space must \textit{vanish} by a simple symmetry argument: On an ensemble-\textit{averaged} level, 
the non-Markovian system has no preferred direction. In other words, the ensemble average of Eq.~(\ref{eq:LE-X0}) is completely symmetric w.r.t. a coordinate inversion $x_0 \to -x_0$. Consequentially, the probability current cannot have any direction. Thus, naively repeating the analysis from Sec.~\ref{SEC:DB}, the system would always \textit{appear} to be in equilibrium. This is, however, not true, as we see by instead considering the FDR~\cite{Kubo1966}, which describes a balance between the \textit{friction kernel} $\gamma$ and thermal noise $\mu$
\begin{equation}\label{eq:FDT}
\langle \mu(t)\mu(s) \rangle  =  k_\mathrm{B} \mathcal{T}_0 \, \gamma(|t-s|).
\end{equation}
As well known for, e.g., viscoelastic fluids, the validity of a FDR
would imply that the system equilibrates in the absence of external driving~\cite{Kubo1966,Maes2013}. 

To check~(\ref{eq:FDT}) for the present model, we rewrite~(\ref{eq:LE-X0}) in the form of a generalized LE 
by converting the time-integral with $K$ in (\ref{eq:LE-X0}) via partial integration into a friction-like integral that involves the ``velocity" $\dot{X}_0$ and the friction kernel $\gamma(|t-s|)$. This yields
\begin{align} 
\int\limits_{0}^{t} \gamma(|t-s|) \dot{X}_{0}(s)\,\mathrm{d}s=& a_{00}X_0 + f_0+ \widetilde{K}(0) {X}_{0}(t) + \widetilde{K}(t) {X}_{0}(0) +\mu(t),
\end{align} 
involving {the noise} $\mu(t)= \xi_0(t) +  \nu (t)$, the integrated kernel $\widetilde{K}$, and the friction kernel $
\gamma(t-s)=  2\gamma_0 \, \delta(t-s) + {\widetilde{K}(t-s)} .
$  
For the case $n=1$, the integrated kernel reads $\widetilde{K}(T)=  (a_{01}a_{10}/a_{11}) e^{a_{11} T/\gamma_1}$.
It can easily be verified [using (\ref{eq:ExampleI}) for the noise correlations] that the FDR holds {only} if
\begin{align} \label{eq:Fulfill-FDR}
a_{10} \mathcal{T}_0 = a_{01} \mathcal{T}_1,
\end{align}which agrees with the DB condition~(\ref{eq:Fulfill-DB}).
Thus, the non-Markovian process is out of equilibrium unless~(\ref{eq:Fulfill-FDR}) holds. This is, for example, never the case for the active microswimmer (where $a_{10}=0$). 
For our $n\,=\,2$ controller (\ref{K_noise_II}), 
we find $\widetilde{K}(T)=k\,\left( 1+ {T}/{\tau} \right)e^{-{T/\tau}}$, and FDR thus amounts to
\begin{align} 
\mathcal{T}_0 k\,\left( 1+ {T}/{\tau} \right)e^{-{T/\tau}}= \mathcal{T}_1 k\,(\tau k  /2\gamma_1)  
\left( 
3  +
%p=1 l=1
T/\tau  \right) \,e^{- T /\tau}.
\end{align} 
There is no pair of $k$, $\tau$ that simultaneously obeys
$\mathcal{T}_0 2\gamma_1 = \mathcal{T}_1 3 \tau k  $ {and} $
% \\
\mathcal{T}_0 2\gamma_1 = \mathcal{T}_1 \tau k $, which would be necessary to fulfill FDR. Thus, in this case, FDR (and DB) are never fulfilled (except for the trivial cases, where $k$ or $\tau$ nullify, or tend towards $\infty$).
For other coupling schemes and $n>1$, we observe that a non-reciprocal system may fulfill FDR, but violate DB. 
We will present a detailed investigation, which is beyond the scope of this paper, in~\cite{Doerries2020}.

In this section, we have seen that non-reciprocity implies an intriguing property of the corresponding non-Markovian stochastic process, i.e., the existence of \textit{nonequilibrium steady states with zero probability currents}. This, in turn, also implies the absence of global particle transport, thus, intrinsic nonequilibrium.
Such states have been considered, e.g., in \cite{Roldan2010}. They commonly occur in active systems~\cite{Zuckermann2015,Korosec2018,Cates2012,Reinken2020}, but can also be found in feedback-controlled systems~\cite{Loos2019}. The reason is that, in both cases, the ``driving'' occurs directly on the level of the stochastic trajectories, yielding, e.g., persistence, but it does not come in the form of a global gradient, i.e., there is no global symmetry breaking (using the language of control theory, one might say that the driving is in a ``closed-loop'' form~\cite{Bechhoefer2005,Gernert2016,Loos2014}). In particular, in the present case, the driving is hidden in the coupling forces. To further investigate this, we will next reconsider the system from an energetic perspective.
\section{Energy $\&$ Entropy}\label{sec:Energy}
\begin{figure}
	\includegraphics[width=0.47\textwidth]{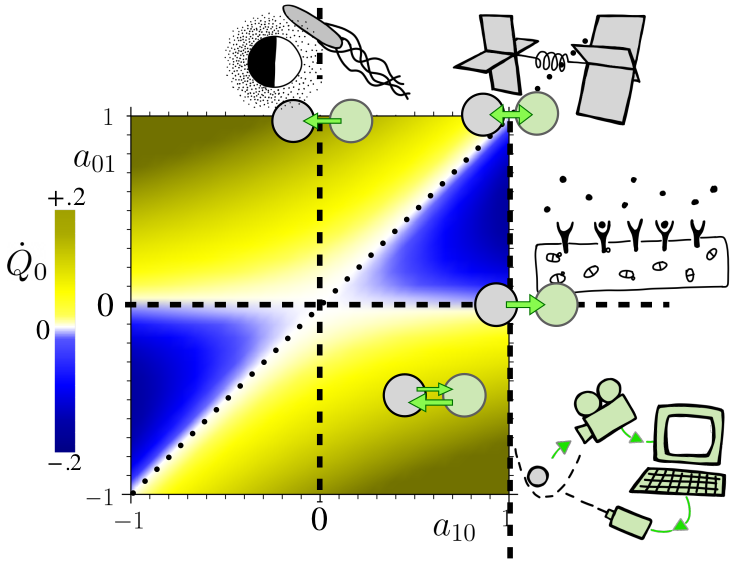}
	\includegraphics[width=0.49\textwidth]{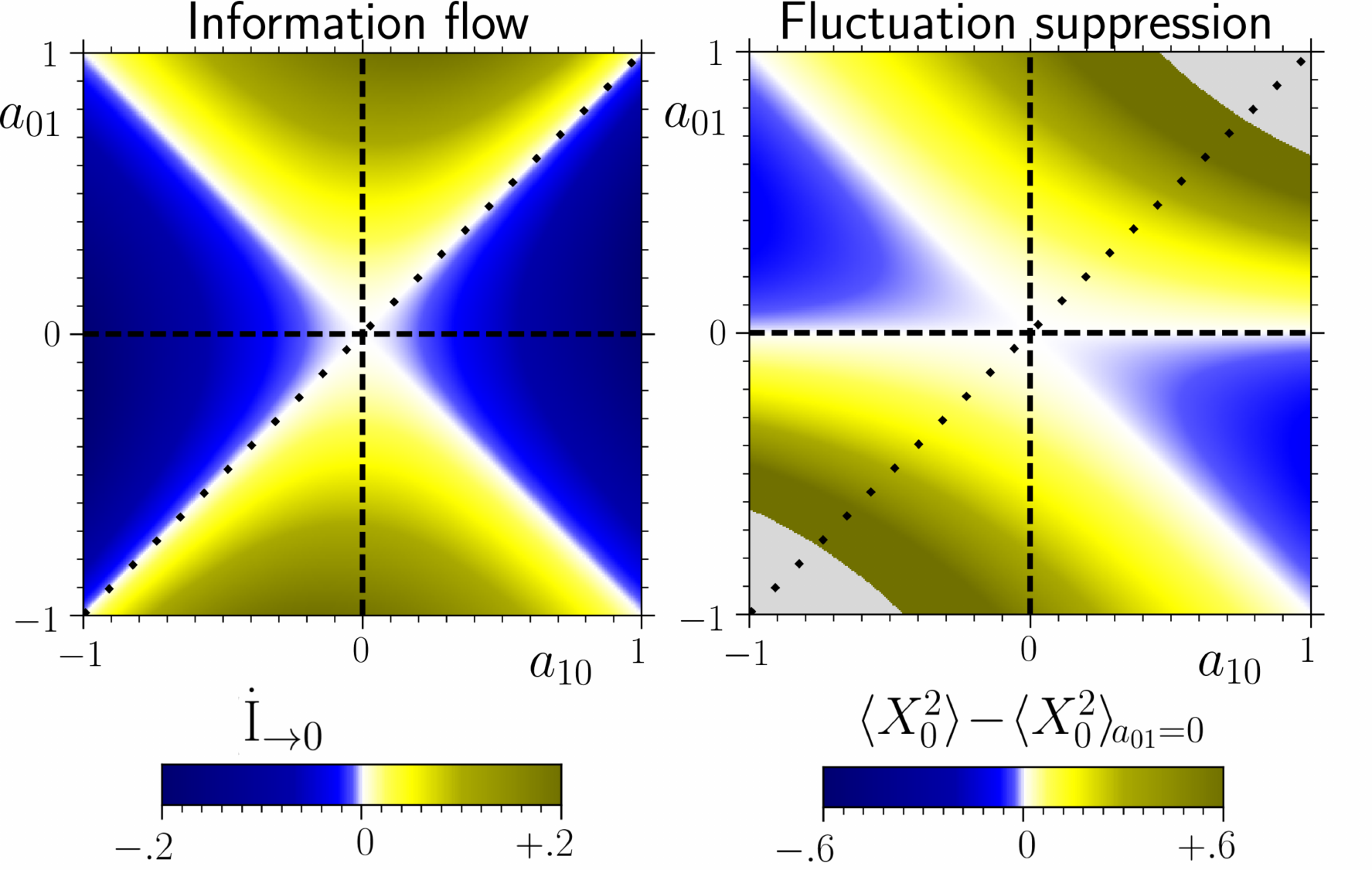} %{\figpath info_suppr.pdf}
	%{FIG_HeatX0.pdf}%{Figures/Frame-active_heat0-crop.pdf}%
	\caption{\textit{Left: }Steady-state mean heat flow $\dot{Q}_0$~(\ref{eq:Q0_n=1}) for (\ref{eq:ExampleI}) at $n=1$. 
		Along the diagonal $a_{01}\!=\!a_{10}$, the super-system is reciprocally coupled, and $X_{0,1}$ may model the angles of vanes coupled by a spring~\cite{Sekimoto2010}. For unidirectional coupling $a_{01}=0$, $X_1$ correspond to a cellular sensor~(\ref{eq:Sensor}) measuring the d.o.f. $X_0$. Along the other unidirectional coupling line $a_{10}=0$, $X_0$ corresponds to the position of an active swimmer in the AOUP model~(\ref{eq:AOUP}). [For further details, see in the main text above and below Eqs.~(\ref{eq:AOUP},\,\ref{eq:Sensor}).]
		For bidirectional non-reciprocal coupling, $X_1$ may model a feedback controller acting on a colloid at $X_0$ [the model~(\ref{eq:Controller_n=1}) with $k=a_{01}$, $\tau=1$ lies on the line $a_{10} = 1$].
		%For bidirectional non-reciprocal coupling, $X_1$ may model a feedback controller~(\ref{eq:Controller_n=1}).
		Here and in the following plots, $a_{11} = a_{00} =-1$, $k_B$ and all other parameters are set to unity.
		\textit{Center}: Information flow to $X_0$. 
		%Blue areas indicate that $X_1$ ``knows'' more about $X_0$ than vice versa corresponding to feedback control regimes ($|a_{10}|>|a_{01}|$). 
		\textit{Right:} Thermal fluctuations of $X_0$ measured by the second moment compared to the uncoupled case ($a_{01}=0$), $\langle X_0^2\rangle -\langle X_0^2\rangle_{a_{01}=0}$. %Blue areas indicate {thermal fluctuation suppression}.
		The grey areas % (upper--right and lower--left corners) 
		indicate unstable regions (where $\langle X_0^2\rangle \to \infty$).
	} %\label{fig:info}
	\label{fig:heat}
\end{figure}
%%%%%%%%%%%%%%%%%%%%%%%%%%%%%
To further unravel the nature of the intrinsic non-equilibrium, we consider the energy flows. 
Sekimoto's framework~\cite{Sekimoto2010} tells us that the fluctuating heat exchange between each $X_j$ and its heat bath along a stochastic trajectory of length $\mathrm d t$ is given by
\begin{align}\label{def:heat}
\delta q_j(t) = [\gamma_j \dot{X}_j(t) - \xi_j(t)]\circ \mathrm{d}{X}_j(t) ,
\end{align}
yielding for the entire super-system a total dissipated energy of $\delta q = \sum_{j=0}^{n} \delta q_j$. Here, $\circ$ indicates Stratonovich calculus. Note that we employ the sign convention that a positive heat flow corresponds to energy flowing from the particle to the heat bath, different from~\cite{Sekimoto2010}.
% % % % % % % % % % % % % % % % % % % % % % % % % %
Using the LE~(\ref{eq:Network}), we can write the ensemble average of the heat rate, denoting $\dot{Q} = \langle \delta q  / \mathrm{d} t\rangle$, $\dot{Q}_j = \langle \delta q_j  / \mathrm{d} t\rangle$, as
\begin{align}\label{eq:HeatZwischenschritt}
\dot{Q}(t)  &=  \sum_{j=0}^{n} \dot{Q}_j(t) = \langle \sum_{j=0}^{n}  (\gamma_j \dot{X}_j(t) - \xi_j(t))\circ \dot{X}_j(t) \rangle 
%
% =\langle \sum_{j=0}^{n}  (a_{jj}X_j(t) + \sum_{i \neq j } a_{ji} X_i(t))\circ \dot{X}_j(t) \rangle 
% \nn
%&=  
=\sum_{j=0}^{n}   a_{jj} \langle X_j (t) \dot{X}_j(t)  \rangle + \sum_{i \neq j } a_{ji} \langle X_i(t)   \dot{X}_j(t)  \rangle 
\end{align}
[recall that $f_0 = 0$]. Now we utilize the steady-state identity $ \langle X_k  \dot{X}_l \rangle =- \langle X_l  \dot{X}_k \rangle~\forall\, k,l$, which readily follows from the fact that the correlations $\langle X_k(t)  {X}_l(t) \rangle$ are time-independent and thus $\frac{\mathrm{d}}{\mathrm{d}t}\langle X_k(t)  {X}_l(t) \rangle = \langle  \dot{X}_k(t) {X}_l(t) \rangle+\langle X_k(t)  \dot{X}_l(t) \rangle =0$. Using these identities, we immediately obtain from~(\ref{eq:HeatZwischenschritt}),
\begin{align}\label{eq:Einput}
\dot{Q}  
 &=   \sum_{j=0}^{n-1}   \sum_{i > j }^{n} (a_{ji}-a_{ij}) \langle X_i   \dot{X}_j  \rangle \geq 0.
\end{align}
Accordingly, if all couplings are reciprocal, the rate of total dissipated energy $\dot{Q}$ is zero, as expected. %
Equation~(\ref{eq:Einput}) further reveals that, in contrast, a non-reciprocal interaction $a_{ij}=a_{ji}$ leads to a net dissipation. Let us discuss this in more depth.

First, we realize that $\dot{Q}$ is nonnegative, as follows from the connection to the total entropy production rate (EP)~\cite{Seifert2012}
\begin{align}\label{eq:Stot_general}
\dot{S}_\mathrm{tot}= 
\sum_{j=0}^{n}\dot{Q}_j/\mathcal{T}_j + \dot{S}_\mathrm{sh} \geq 0,
\end{align}
with $S_\mathrm{sh}$ being the ensemble average of the fluctuating {multivariate} (joint) Shannon entropy 
$s_\mathrm{sh}=-k_\mathrm{B}\ln[ \rho_{n+1}(\underline{x})]$, and $\dot{S}_\mathrm{sh} \equiv 0$ in steady states. Noteworthy, (\ref{eq:Stot_general}) describes the actual total thermodynamic EP only when all d.o.f. have a physical interpretation. In other cases its meaning is debatable. However, in any case, the second law $\dot{S}_\mathrm{tot}\geq 0$ holds [as formally shown below in~(\ref{eq:Sj1})], where $\dot{S}_\mathrm{tot}= 0$ in thermal equilibrium.

Second, according to the first law of thermodynamics, $\delta q =  \delta w + \mathrm{d} u $, the net dissipation associated with each non-reciprocal interaction~(\ref{eq:Einput}), must result from work $\langle \delta{w} \rangle$ applied to the system, while the internal energy is conserved in steady states, $\langle \mathrm{d}{u} \rangle=0$. In other words, the total entropy production is due to a \textit{positive energy input} at rate $\dot{W}=\dot{Q}\geq 0$~(\ref{eq:Einput}) into the system. 
{Where does this energy come from? Because fundamental physical interactions are generally reciprocal, in order to realize a non-reciprocal coupling some (external) mechanism is necessary, which is here not explicitly modeled but ``hidden'' in the equations within the non-reciprocity. 
The positive energy input $\dot{W}=\dot{Q}\geq 0$~(\ref{eq:Einput}) gives the minimal energy needed (by this mechanism) to sustain the non-reciprocal coupling.}
We also note that a positive energy input on the level of the fluctuating trajectories is considered a defining property of active systems~\cite{ramaswamy2010mechanics,ramaswamy2017active,nardini2017entropy,fodor2016far,dauchot2019chemical}. As we see here, it can be introduced in the form of a non-reciprocal interaction.

%
%

%and a contribution from the additional d.o.f. $X_{j>0}$.
Next, we take a closer look at the individual heat flow between $X_0$ and its bath. We focus on $\dot{Q}_0$, as it is a characteristic thermodynamic quantity and it is independent of whether all d.o.f. have a clear physical interpretation, or not, and independent of the employed description (Markovian or non-Markovian). To calculate the steady-state ensemble average, we again utilize $\langle X_j \dot{X}_j\rangle = 0$ and $\langle X_l \xi_j\rangle = 0$ for all $j\neq l$, and therewith find from (\ref{def:heat}) directly
\begin{align}\label{eq:Q0_general}
\dot{Q}_0 = \sum_{j> 0}^{n} a_{0j}  \langle  X_j \dot{X}_0  \rangle =
\sum_{i=0}^{n}\sum_{j>0}^{n} \frac{a_{0j}a_{0i}}{\gamma_0} \langle  X_j X_i  \rangle  =\dot{W}_0.
%\dot{Q}_0 = \sum_{j>0}[ a_{0j}^2 \langle  X_j^2 \rangle +  a_{0j} \,a_{00} \langle  X_0 X_j  \rangle ] . 
\end{align}
Likewise, one can calculate the heat flows of the other d.o.f.
$\dot{Q}_l =\sum_{i=0}^{n}\sum_{j\neq l} \frac{a_{lj}a_{li}}{\gamma_l} \langle  X_j X_i  \rangle $. 
It should be noted that by writing down this expression for the dissipation of $X_{j>0}$ and the total EP~(\ref{eq:Stot_general}), we implicitly assume that all $X_{j}$ are \textit{even} under time-reversal, that means, position-like variables. In contrast, odd variables would not contribute to the total EP, see~\cite{Shankar2018}.
We note that for active matter models the parity of $X_{j>0}$ is in fact a nontrivial aspect, and subject of an ongoing debate, see e. g., \cite{Shankar2018,Caprini2019,Holubec2020,Pietzonka2019}.

Together with the correlations $\langle X_iX_j\rangle$ that are derived in Appendix~\ref{sec:analytical_solutions}, Eqs.~(\ref{eq:Stot_general},\,\ref{eq:Q0_general}) represent analytical expressions for heat flow and entropy production for any $n$. 
For example,
in the case $n=1$ (which was also discussed in~\cite{Crisanti2012}), where the expression significantly simplifies, we find from (\ref{eq:Q0_general},\,\ref{eq:Stot_general}) in combination with~(\ref{eq:Corr_n=1})
\begin{align}
\dot{S}_\mathrm{tot} & =  \frac{k_\mathrm{B}}{\mathcal{T}_0\mathcal{T}_1} \frac{( a_{10}\mathcal{T}_0  - a_{01} \mathcal{T}_1)^2}{-a_{00}\gamma_1 -a_{11}\gamma_0} \geq 0,\label{eq:Stot_n=1} \\
\dot{Q}_\mathrm{0} & =k_\mathrm{B} \frac{a_{01}(a_{10}\mathcal{T}_0  - a_{01} \mathcal{T}_1)}{(a_{00}\gamma_1 +a_{11}\gamma_0)}. \label{eq:Q0_n=1}
\end{align}

From~(\ref{eq:Stot_n=1}) one immediately sees that the EP vanishes if, and only if, DB~(\ref{eq:Fulfill-DB}) and FDR~(\ref{eq:Fulfill-FDR}) are fulfilled (as one shall expect). Thus, all three notions of equilibrium are consistent. Further, if~(\ref{eq:Fulfill-DB}) is fulfilled, also the heat flow vanishes. As we show in Sec.~\ref{sec:underdampedCases}, this also holds for the corresponding underdamped model, see Eq.~(\ref{eq:underdampedHeat}).
{Thus, now we have convinced ourselves that the non-reciprocal systems which are simultaneously coupled to baths at different temperatures can really reach states of thermal equilibrium, if~(\ref{eq:Fulfill-DB}) holds.}

Let us now take a closer look at the heat flow~(\ref{eq:Q0_n=1}) for different coupling schemes, shown in Fig.~\ref{fig:heat} for %isothermal conditions, i.e., 
$\mathcal{T}_{0}=\mathcal{T}_1$, $a_{00}=a_{11}$ and $n=1$. Note that these isothermal conditions allow to better investigate the effect of non-reciprocity and, at the same time, are most realistic in regard to experimental realizations.
For example, this could represent a system of two colloidal particles trapped in a harmonic potential of stiffness $a_{00}=a_{11}$ and coupled with each other with the help of an external setup similar to~\cite{Khadka2018,Lavergne2019}.
%. Since we are primarily interested in the role of non-reciprocal coupling, we focus here and in the remainder of this paper on isothermal conditions, i.e., $\mathcal{T}_{j}=\mathcal{T}$ for all $j>0$. 
When the system is reciprocally coupled (along the dotted diagonal), it equilbrates and the heat flow nullifies. Then, the EP~(\ref{eq:Stot_n=1}) is zero as well. The heat flow also vanishes in the trivial case $a_{01}=0$, i.e., when $X_0$ does not ``see'' $X_1$ (dashed horizontal line), as is the case when $X_1$ corresponds to a sensor~\cite{hartich2016sensory}. As one would expect, being measured does not bring $X_0$ out of equilibrium. If the unidirectional coupling is reversed ($a_{10}=0$), the heat flow is strictly nonnegative (dashed vertical line). This suits to the idea that $X_0$ is an active swimmer: the swimmer eventually heats up the surrounding fluid, but never has a net cooling effect. Remarkably, for cases with bidirectional non-reciprocal coupling, we observe that, $\dot{Q}_0$ can also become \textit{negative}. We will discuss this in more depth in the next section.

\subsection{Conditions for negative heat flow}\label{SEC:revHeatFlow}
When $\dot{Q}_0$ is negative (as in the blue regions of Fig.~\ref{fig:heat}), heat is constantly flowing 
out of the bath (on average).This happens due to the coupling with another (or multiple) subsystem(s) $X_1$, although the other subsystem is \textit{not} colder, which would be a trivial case of heat extraction. 
Let us take a moment and think about the meaning of this observation. We aim to remind the reader that a steady-state heat flow induced by an non-conservative external force (e.g., a constant, a time-dependent, or a space-dependent driving force) acting on a passive, Markovian system is strictly {nonnegative}, as dictated by the second law, $\dot{Q}_0/ \mathcal{T}_0 = \dot{S}_\mathrm{tot}\geq 0$. %
Loosely speaking: ``Stirring a particle from outside will eventually heat up the environment.''
Here we find that, in contrast, the non-conservative force $a_{01} X_1$, or $F_\mathrm{c}$ in the notation (\ref{eq:LE-X0_Feedback}), \textit{can} induce a negative heat flow, $\dot{Q}_0 < 0$.
Thus, $F_\mathrm{c}$ can be viewed as an external force, which stirs the particle in a clever way, and thereby cools down the particle's environment.
 (The underlying reason is the usage of extracted information, see Sec.~\ref{sec:Info}.) The negative sign of $\dot{Q}_0$ implies a steady extraction of energy from the bath, which is converted into work $\dot{W}_0$, i.e., a (potentially useful) form of energy. %
It is, of course, well-known that such an energy extraction can be realized by ``Maxwell-demon''-type of devices~\cite{Koski2014,maxwell2001theory}. Here we see that the non-reciprocally coupled d.o.f. represents a minimal, time-continuous version of such a device, where the control action is automatically encoded in the non-reciprocity of the coupling. 
Note that the total EP, which is proportional to the sum over both heat flows, $\dot{Q}_0+\dot{Q}_1$, is strictly positive also in this case, i.e., the isothermal ``Maxwell demon'' $X_1$ must heat up its own environment. 

%
%
%
%
% % % %%%%%%%%%%%
\begin{figure}
	\begin{minipage}[t]{0.49\textwidth}
		%\textcolor{white}{.}\vspace{-0.3cm}\\
		\begin{flushleft} (a) Heat flow\end{flushleft}
	\end{minipage}
	\begin{minipage}[t]{0.49\textwidth}
		\begin{flushleft} (b) Information flow\end{flushleft}
	\end{minipage}
	\begin{minipage}{0.49\textwidth}
		\includegraphics[width=0.93\textwidth]{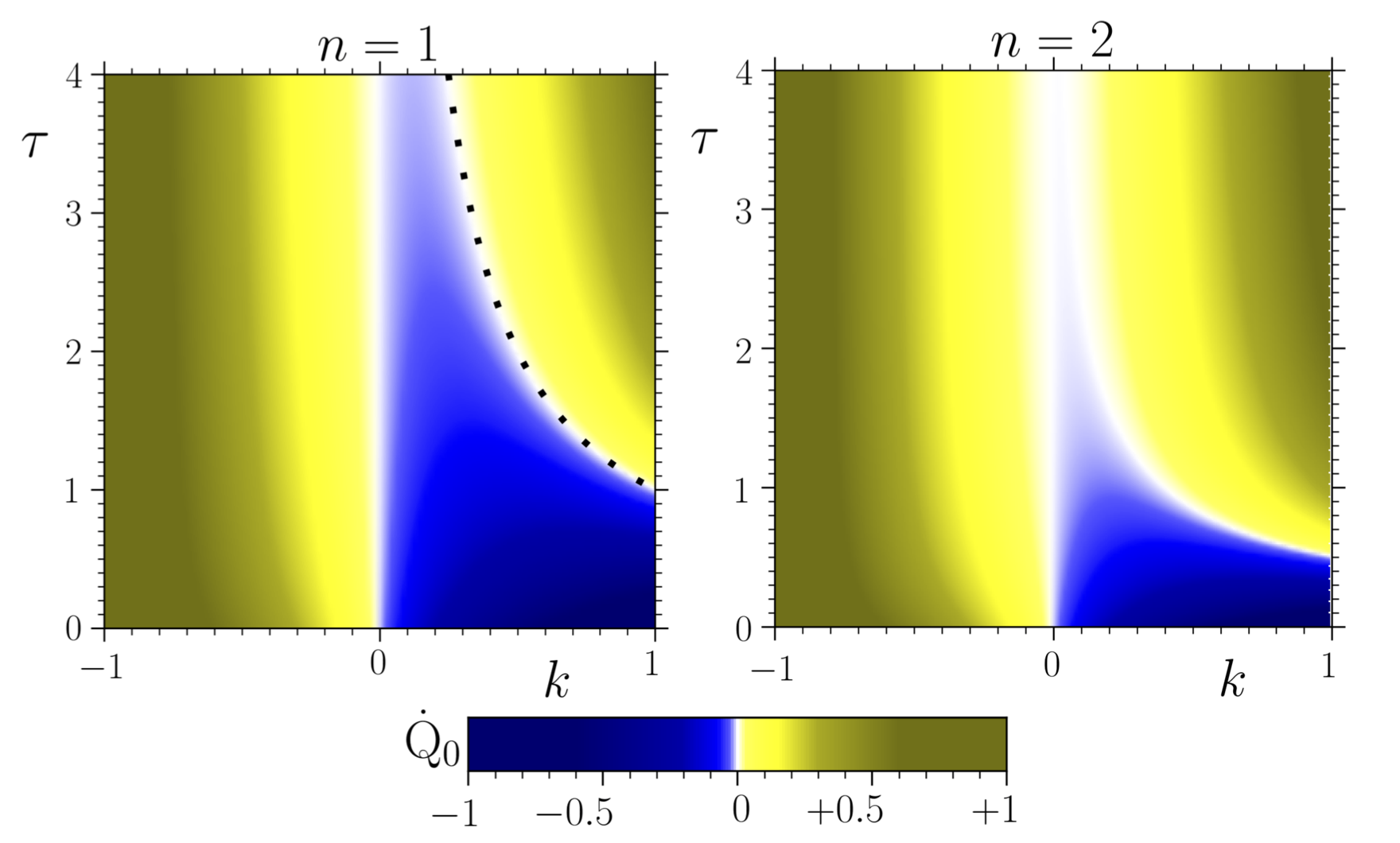}\\
	\end{minipage}
	\begin{minipage}{0.49\textwidth}
		\includegraphics[width=0.99\textwidth]{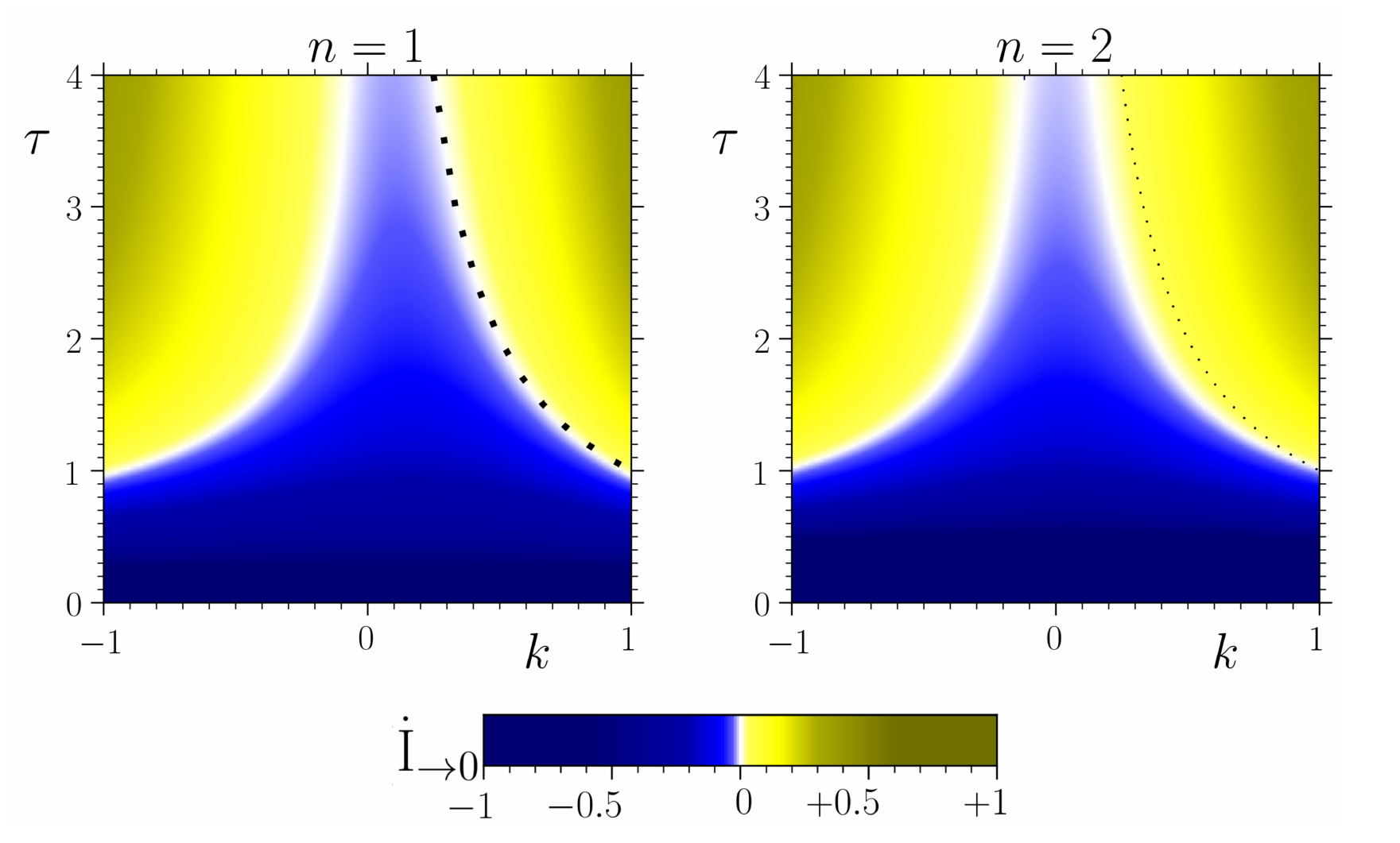} \\
	\end{minipage}
	\caption{
		(a) Mean heat flow $\dot{Q}_0$ from $X_0$ to its bath~(\ref{eq:Q0_general}, \ref{eq:Q0_n=1}), and (b) information flow $\dot{I}_{\to 0}$ from all other d.o.f. to $X_0$~(\ref{eq:info_n=2}); in the controller models with $n=1$ (\ref{eq:Controller_n=1}), and $n=2$ (\ref{K_noise_II}), vs. feedback gain $k$ and time delay $\tau$. {Further, $a_{00}=-1$ and} all other parameters are set to unity.	
		For $n=1$, DB and FDR~(\ref{eq:Fulfill-DB}) hold at $k= 1/\tau$ (dotted lines), i.e., the system is in equilibrium. 
		We have added a corresponding thin line also in the $n=2$ plot of the information flow, serving as a guide to the eye.
		Trivially, in the uncoupled case ($k=0$), the subsystem $X_0$ is equilibrium as well (for arbitrary $n$). 
		Note that the $(n=1)$-controller with $\tau=1$, corresponds to the system in Fig.~\ref{fig:heat} along the (dashed) line $a_{10}=1$ with $a_{01}=k$. 
		%}\label{fig:info_n=1-2}
	}\label{fig:heat_n=1-2}

\end{figure}
%%%%%%%%%%%%%%%%%
%
%
%
% 
%
%\subsection{\remove{Conditions for reversed heat flow}}%\label{SEC:revHeatFlow}
To find out under which conditions the negative heat flow occurs for $n=1$ and $2$ [with the parameter setting from~(\ref{K_noise_II},\,\ref{eq:Controller_n=1})], we vary the two important parameters, the feedback gain $k$ and delay time $\tau$. 
Figure~\ref{fig:heat_n=1-2} reveals that the heat flow $\dot{Q}_0$ is qualitatively and quantitatively similar for $n=1$ and $2$. The similarity of the two cases is indeed striking, given the differences between both systems. In particular, we here compare systems with monotonic memory kernel $K(t=t')$, vs. non-monotonic $K(t=t')$ which nullifies at $t=t'$ (for $n=2$). At $n=1$, the feedback force $k\int K(t-t')X_0(t')\mathrm{d}t'$ mostly depends on the instantaneous position $X_0(t)$, while at $n=2$ it is \textit{independent} of the latter, and mostly depends on $t-\tau$. Further, in regard to the Markovian super-system, there is a direct coupling from $X_1$ to $X_0$ in the case $n=1$, while this coupling is only indirect (via a third sub-system) in the case $n=2$. 
Nevertheless, the (blue) area of reversed heat flow lies in the same region of the $(\tau,k)$-plane and is of similar size. 
Also, in both cases, it only occurs if $k>0$.

In the context of control theory, it is common to characterize feedback loops as positive or negative feedback, according to the question whether a small perturbation (from the desired state) is enhanced, or reduced by the feedback~\cite{Bechhoefer2005}. In the present case, $ k<0$ corresponds to positive feedback, while $k>0$ is negative feedback, see App.~\ref{APP:Feedback} for an explanation on the terminology and an illustration. Thus, in both models, only {negative feedback} may induce a negative heat flow.

Besides the trivial case, $k=0$, there is, for both $n$, a second line in Fig.~\ref{fig:heat_n=1-2} along which the heat flow vanishes. For $n=1$, this line corresponds to parameters where DB and the FDR~(\ref{eq:Fulfill-DB}) are fulfilled (dashed line), i.e., the system is in equilibrium. For $n=2$, DB and the FDR are generally broken for all $(\tau,k)$. This second line hence reveals another interesting property of non-Markovian systems: \textit{They may be out of equilibrium without exhibiting dissipation (zero heat flow)}, in sharp contrast to reciprocal systems. In our system, such a state is found for $n>1$ and non-reciprocal coupling only. From the viewpoint of the non-Markovian process $X_0$ this is indeed a bit puzzling. If $X_0$ is in a true nonequilibrium steady-state, there must be an associated entropy production. However, the zero heat flow indicates zero medium entropy production. Thus, where does the entropy go? To answer this question, we shall consider the entropy balance of the individual subsystem $X_0$, as we will do the next section.

We note that a NESS with zero heat flow and regimes of negative heat flow may also occur in systems with $\delta$-distributed memory, which are, moreover, nonlinear, as we have reported in~\cite{Loos2019}. Further, such states may also occur in Markovian (reciprocal) systems with non-Gaussian noise. As was shown in \cite{Kanazawa2013}, the presence of nonlinear forces is then a necessary condition, different from the reversed heat flow induced by non-reciprocal coupling or time-delayed feedback.
\section{Information} \label{sec:Info}
Now we turn to an information-theoretical investigation of non-reciprocal coupling. The motivation of this is two-fold. First, it will help us better understand the previous observations, for example: Why is heat extraction only possible 
for negative feedback (see Fig.~\ref{fig:heat_n=1-2}), and only 
if $|a_{10}|>|a_{01}|$ (Fig.~\ref{fig:heat})? Until now, these conditions seem arbitrary. %By the information-theoretical analysis, we will obtain a deeper understanding.
Second, by also considering information flows, we will be able to describe the entropy balance of an \textit{individual} subsystem whereas, so far, we have studied entropic properties of the entire super-system only. This is especially important in situations where only one part of the system is observable (or has a direct physical interpretation).
%
%%%%%%%%%%%%%%%%%%

It has been established in previous literature~\cite{allahverdyan2009thermodynamic,hartich2016sensory,horowitz2014second} that the entropy flow associated with the exchanged information between two coupled subsystems (say $X_0$ and $X_1$), is associated with the \textit{information flows} between them. This quantity is closely connected to the mutual information~\cite{horowitz2014second,Dabelow2019}, which describes the total amount of information exchange in the entire supersystem, but is, in contrast to the information flows, not directed. While information flows are already common to investigate discrete systems~\cite{Horowitz2014,horowitz2014second,Ito2015,Koski2014}, this quantity is less established for time- and space-continuous systems (which time-continuous feedback)~\cite{Horowitz2014,Horowitz2010}. First steps in this direction have been undertaken in~\cite{hartich2016sensory,horowitz2014second} and in~\cite{allahverdyan2009thermodynamic} (where the reciprocally coupled $n=1$ case was studied). It should be noted that there
are various other notions of information 
flows and information exchanges, which are more appropriate in other contexts, see~\cite{Horowitz2014} for an
educational overview. 

However, the previously developed framework based on information flows, and the definition of the mutual information itself, are only applicable for situations where  \textit{two} subsystems exchange information ($n=1$). Here we will generalize this framework to arbitrary system sizes and topologies. 

We start by considering the total temporal derivative of the Shannon entropy~(\ref{eq:Stot_general}), i.e., 
\begin{align}\label{eq:SsysTotal}
 \frac{\dot{s}_\mathrm{sh}}{k_\mathrm{B}}=& \frac{-\partial_t \rho_{n+1} }{\rho_{n+1}}+  \sum_{j=0}^{n} \frac{-\left(\partial_{x_j} \rho_{n+1}\right)\dot{X}_j }{ \rho_{n+1}} . 
%\\
%
\end{align} 
In steady states, the first term naturally vanishes. To calculate the ensemble average of the second term, 
we use $\langle \dot{X}_j A(x_j,t) \rangle = \int J_j A(x_j,t) \mathrm{d}x_j$~\cite{Reimann2002,Seifert2012}, with the probability currents $J_j$. 
%$J_j= \left[\frac{F_j}{\gamma_j} - \frac{ k_\mathrm{B}\mathcal{T}_j}{\gamma_j }\, \partial_{x_j}\right]\rho_{n +1}$. 
We consider natural boundary conditions $\lim_{x\to \pm \infty}\rho(\underline{x}) =0$, and denote improper integrals $\lim_{r\to\infty} \int_{-r}^{r}$ simply as $\int$. With these tools, we find the ensemble average of each summand of~(\ref{eq:SsysTotal})
\begin{align}
%\begin{split}
%\dot{S}_\mathrm{sh}= &\sum_{j=0}^{n} 
\left \langle  \frac{-\left(\partial_{x_j} \rho_{n+1}\right)\dot{X}_j }{\rho_{n+1}} \right \rangle = &  
 \int  \frac{-\left(\partial_{x_j} \rho_{n+1}\right) J_j }{  \rho_{n+1}} \, \mathrm{d}\underline{x} 
%\nn
%=&
= -\int \underbrace{\left[ \ln(\rho_{n+1}) J_j\right]_{-\infty}^{\infty}}_{\to 0} \,\mathrm{d}\underline{x}_{i\neq j} 
+ \int  \ln \rho_{n+1}(\underline{x})  \partial_{x_j} J_j   \,\mathrm{d}\underline{x} 
\nn
=&-\underbrace{ \int \! \ln \frac{\rho_{1}(x_j)}{  \rho_{n+1}(\underline{x})}\partial_{x_j} J_j \,\mathrm{d}\underline{x} }_{ =  \dot{I}_{\to j} } + \int \ln\rho_1(x_j) \,\partial_{x_j} J_j  \,\mathrm{d}\underline{x} ,
\label{eq:Iflow}
\end{align}
where we have introduced the multivariate information flow ${\dot{I}}_{\to j}$ to $X_j$ (we note that when applied to the case $n=1$, the here defined ${\dot{I}}_{\to j}$ reduces to the information flow from~\cite{horowitz2014second,allahverdyan2009thermodynamic}, with the sign convention as in \cite{allahverdyan2009thermodynamic}.
We stress that the involved information flow is from \textit{all} other d.o.f. $\{X_{l\neq j}\}$ to $X_j$. Even if not directly coupled with each other, two d.o.f. can exchange information through a third d.o.f..
Furthermore, we recall that 
thermal equilibrium is characterized by vanishing probability current. Thus, from the definition~(\ref{eq:Iflow}), one can see that in equilibrium all individual information flows are necessarily zero.
%, whereas 
 
To further proceed, we utilize the closed, {multivariate} Fokker-Planck equation (\ref{eq:FPE-vector}), and find
\begin{align}
...=&- {\dot{I}}_{\to j} + \int \ln\rho_1(x_j)\, \left\{-\partial_{t}\rho_{n+1} +\sum_{i\neq j}\partial_{x_i} J_i  \right\} \,\mathrm{d}\underline{x} 
%\nn
=
- {\dot{I}}_{\to j} + \frac{\dot{S}^j_\mathrm{sh}}{k_\mathrm{B}}-  \sum_{i\neq j}\int \ln \rho_1(x_j) \, \underbrace{\left[  J_i \right]_{-\infty}^{\infty}}_{\to 0} \,\mathrm{d}\underline{x}_{l\neq i},
\end{align}
where we have introduced the change of the Shannon entropy of the \textit{marginal} pdf $\rho_{1}(x_j)$
\begin{align}
\dot{S}^j_\mathrm{sh}=-k_\mathrm{B}\int \!  \ln   \rho_{1}(x_j) \partial_{t} \rho_{1}(x_j)\, \mathrm{d}{x_j}.
\end{align}
In sum, we have shown that 
\begin{equation}\label{eq:ShannonSubTotal}
\dot{S}_\mathrm{sh}= -k_\mathrm{B}\sum_{j=0}^{n}  {\dot{I}}_{\to j} + {\dot{S}^j_\mathrm{sh}}{}.
\end{equation}

Let us now consider the multivariate information flows defined in (\ref{eq:Iflow}) in more detail. For the case $n=1$, it has been shown that 
$\dot{I}_{\to 0} +\dot{I}_{\to 1}=\dot{\mathcal I}$~\cite{allahverdyan2009thermodynamic}, i.e., the individual information flows sum up to the temporal derivative of the mutual information. As we show in Appendix~\ref{sec:AppInfo}, for systems with multiple subsystems ($n\geq 1$), the individual information flows sum up to the multivariate generalization of the mutual information
\begin{align}\label{def:mutualInfo}
	\mathcal{{I}}(\underline{x}) =\int \rho_{n+1}(\underline{x}) \ln \frac{  \rho_{n+1}(\underline{x})}{ \rho_{1}(x_0)\rho_{1}(x_1)\dots\rho_{1}(x_n) }  \,\mathrm{d}\underline{x}.
\end{align}
For $n=1$, this reduces to the usual mutual information. Just like the latter, we have, on the one hand, $\sum_{j=0}^{n} \dot{I}_{\to j}=\dot{\mathcal I}$, and, on the other hand, $\dot{\mathcal{{I}}}=0$ in steady states (because the pdfs are time-independent).
(We note that 
	there is not a unique way of generalizing the mutual information to systems with more than two subsystems, see~\cite{Mcgill1954,Te1980,Ting1962,Srinivasa2005}).

Since $\mathcal{\dot{I}} =0$, the information flows among all d.o.f. $X_j$ in total cancel each other out (thus, from an information-theoretical point of view, the super-system as a whole is ``closed''). 
However, they constitute an important contribution to the entropy balance, when an \textit{individual }subsystem is considered.

To see this, we reconsider the summands of~(\ref{eq:SsysTotal}), and rewrite them using the FPE~(\ref{eq:FPE-vector}) as
\begin{align}
\frac{-\left(\partial_{x_j} \rho_{n+1}\right)\dot{X}_j }{k_\mathrm{B}^{-1} \rho_{n+1}}= \underbrace{ \frac{\gamma_j \,J_j( \underline{x},t )\dot{X}_j }{\mathcal{T}_{j}\,\rho_{n+1}} }_{:=k_\mathrm{B}\dot{s}_\mathrm{tot}^j}- \frac{\dot{q}_j }{\mathcal{T}_{j}}  . \label{eq:STotalJ}
\end{align}  
Combining (\ref{eq:SsysTotal}, \ref{eq:Iflow}, \ref{eq:STotalJ}),
we obtain the entropy balance of each subsystem
% shows that a definition of a fluctuating ``total EP per subsystem'', $\dot{s}_\mathrm{tot}^j $,
%is meaningful, as it  has the non-negative ensemble average
\begin{align}
\dot{S}_\mathrm{tot}^j & = \dot{S}^j_\mathrm{sh}  -k_\mathrm{B}{\dot{I}}_{\to j} + \frac{\dot{Q}_j}{\mathcal{T}_j} =
 \int \frac{\gamma_j J_j( \underline{x},t )^2}{\mathcal{T}_{j}\,\rho_{n+1}}  \,\mathrm{d}\underline{x}\geq 0 , \label{eq:Sj1}
\\
\dot{S}_\mathrm{tot}  & \stackrel{(\ref{eq:Stot_general})}{=} \sum_{j=0}^{n} \frac{\dot{Q}_j}{\mathcal{T}_j} + \dot{S}_\mathrm{sh}
\stackrel{(\ref{eq:ShannonSubTotal},\ref{eq:Sj1})}{=}
\sum_{j=0}^{n} \dot{S}_\mathrm{tot}^j   \geq 0 . \label{eq:Sj3}
\end{align}%and gives the entropy balance for each d.o.f..
 With~(\ref{eq:Sj3}) we have recovered the mean total EP~(\ref{eq:Stot_general}).

Further, Eq.~(\ref{eq:Sj1}) may be seen as a generalized second law for each d.o.f., giving the entropy balance of an individual sub-system. In steady states, where $\dot{S}^j_\mathrm{sh}=0$, it implies
\begin{align}\label{eq:GeneralizedSecondLaw}
{\dot{Q}_j} \geq k_\mathrm{B}\mathcal{T}_j\,{\dot{I}}_{\to j},
\end{align}
consistent with~\cite{horowitz2014second,allahverdyan2009thermodynamic}.

\begin{figure}
	\includegraphics[width=0.38\textwidth]{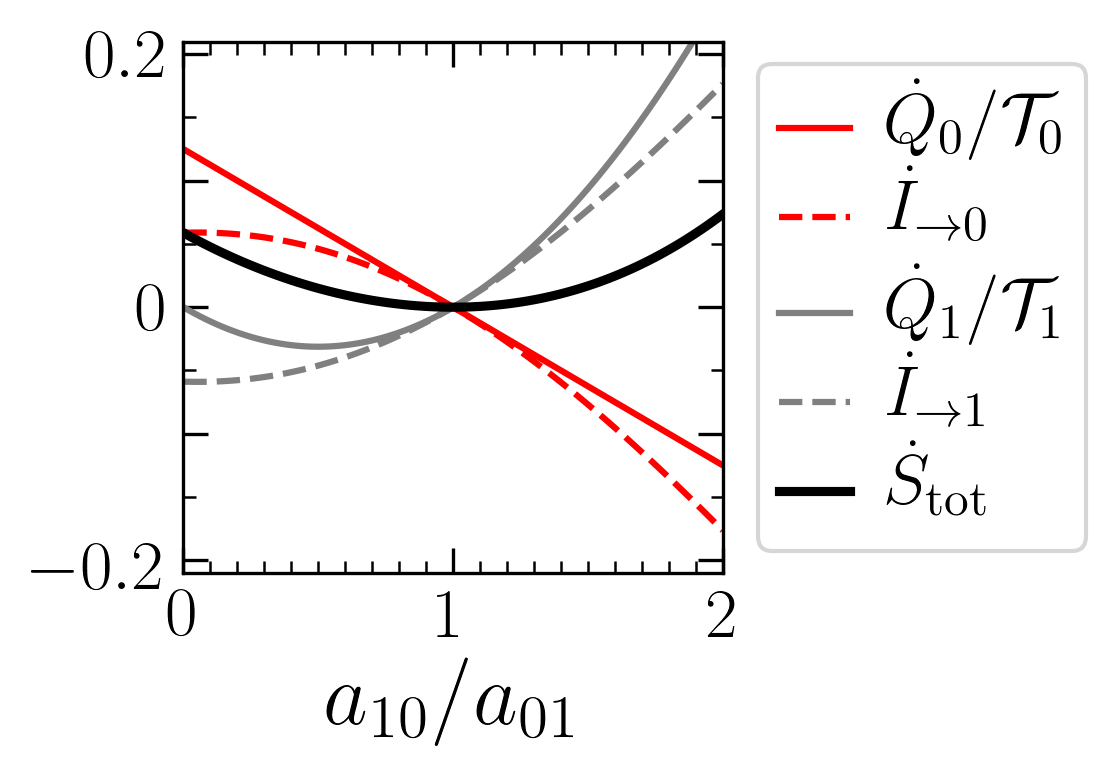} %{\figpath Heat_Info.png} %
	\caption{Information $\dot{I}_{\to j}$~(\ref{eq:Info-flow-0}) and heat $\dot{Q}_{j}$~(\ref{eq:Q0_n=1}) flows vs. $a_{01}/a_{10}$, for $n=1 $. 
		%At $a_{01}=a_{10}$, the system is reciprocally coupled and displays no net heat and information flow, and has zero EP (black line). 
		The heat flow (solid red and grey lines) of each d.o.f. is bounded from below by the information flow (dashed lines), as predicted by
		%the generalized second law
		~(\ref{eq:GeneralizedSecondLaw}). The total EP is given by the sum over $\sum_{j=0,1}\dot{Q}_j -\dot{I}_{\to j}$. The plots pertain to $a_{11} = a_{00} =-1$, and all other parameters and $k_\mathrm{B}$ are set to unity.
	}\label{fig:Info+Heat}
\end{figure}
Equation~(\ref{eq:GeneralizedSecondLaw}) states that a negative steady heat flow, $\dot{Q}_0<0$, is only possible, if $\dot{I}_{\to 0}<0$, i.e., information is flowing from the $X_0$ to the rest of the system. The more information about $X_0$ is gathered by the other $X_{j>0 }$ (the controller d.o.f.), the more heat can be extracted from the bath. Figure~\ref{fig:Info+Heat} shows (for $n=1$) the information and heat flows, as well as the total EP, which are all connected via~(\ref{eq:Sj1}, \ref{eq:Sj3}). It also illustrates that, in the reciprocal and isothermal case, there is no ``entropic cost'' (zero EP) , but, at the same time, no net information extraction is achieved, nor is a heat flow induced.

% % % % % % % % % % % % % % % % % % % % % 
% % % % % % % % % % % % % % % % % %

Due to the linearity of the model, we can calculate the information flows analytically. The steady-state pdfs are multivariate Gaussians with zero mean and with the covariance matrix $(\underline{\underline{\Sigma}})_{i j}=\langle X_i X_j\rangle$, which are described in Appendix~\ref{sec:analytical_solutions}. To derive explicit expressions for the steady-state information flows, it turns out to be most convenient to start with~(\ref{eq:Iflow}). Using the general property of normal distributions, $\partial_{x_j} \rho_{n+1}(\underline{x})= 
-(\underline{\underline{\Sigma}}^{-1}\underline{X})_j \rho_{n+1}(\underline{x}) 
$ [and recalling $\dot{S}^j_\mathrm{sh}=0$], we find 
\begin{align}\label{eq:generalFormula_Infoflow0}
{\dot{I}}_{\to j} 
=&\left \langle \frac{\left(\partial_{x_j} \rho_{n+1}\right)\dot{X}_j }{\rho_{n+1}} \right \rangle
= -\left \langle(\underline{\underline{\Sigma}}^{-1}\underline{X})_j \dot{X}_j \right \rangle.
\end{align}
%
%  $$ -  \sum_{l=0}^{n} (\underline{\underline{\Sigma}}^{-1})_{jl}  \langle {X}_l \dot{X}_j   \rangle $$
%
Inserting the Langevin equations~(\ref{eq:Network}), utilizing $  2 \langle X_l\dot{X_l}\rangle= \mathrm{d}\langle X_l^2\rangle/\mathrm{d}t= 0$
and $\langle X_l \xi_j\rangle = 0$ for $j\neq l$, we obtain the general formula 
\begin{align}\label{eq:generalFormula_Infoflow}
{\dot{I}}_{\to j} &=
-  \sum_{l\neq j}^{n} (\underline{\underline{\Sigma}}^{-1})_{jl}  \langle {X}_l \dot{X}_j   \rangle
%\nn&=
=
- \sum_{i=0}^{n}  \sum_{l\neq j}^{n}   \frac{a_{ji}}{\gamma_j}
\left(\underline{\underline{\Sigma}}^{-1}\right)_{jl} \left(\underline{\underline{\Sigma}}\right)_{li} 
%\langle {X}_l {X}_i \rangle\right] 
=-\frac{a_{jj}}{\gamma_j}
+   \frac{1}{\gamma_j}
\left(\underline{\underline{\Sigma}}^{-1}\right)_{jj} \left(\underline{\underline{a}}\underline{\underline{\Sigma}}\right)_{jj}.
\end{align}
Equation~(\ref{eq:generalFormula_Infoflow}) represents in combination with~(\ref{eq:Solution-Corr}), an analytic expression for the steady-state information flow to any sub-system in (super-)systems of arbitrary sizes.  

\subsection{A single non-reciprocal interaction, n=1}
We are now in the position to clarify the information-thermodynamic implications of non-reciprocal coupling. First we start with $n=1$,
where we find from (\ref{eq:generalFormula_Infoflow}) in combination with (\ref{eq:Corr_n=1}) %severely simplifies. In particular, we identify from equation~(\ref{eq:generalFormula_Infoflow}) the close connection to the heat flow~(\ref{eq:Q0_general}), yielding (if $a_{01}=0$)
\begin{align}\label{eq:Info-flow-0_general}
	{\dot{I}}_{\to 0}
%	&
	=-\frac{a_{00}}{\gamma_0}-
	\frac{
		\langle X_1^2\rangle (a_{00}\langle X_0^2\rangle+a_{01}\langle X_0X_1\rangle)	
	}
	{\gamma_0(\langle X_0X_1\rangle^2 -\langle X_0^2\rangle \langle X_1^2\rangle) }=
    \frac{ [a_{01} \mathcal{T}_1-a_{10} \mathcal{T}_0][ a_{00} a_{01}/\mathcal{T}_0 +
    	a_{11} a_{10} \gamma_0/(\mathcal{T}_1\gamma_1)]
    }
    { \mathcal{T}_0(a_{00} \gamma_1+a_{11}
    	\gamma_0)^2+  \frac{a_{01}^2\mathcal{T}_1}{\gamma_0 \gamma_1}  -2 \mathcal{T}_0 a_{01} a_{10} +a_{10}^2  \frac{\mathcal{T}_0^2}{\mathcal{T}_1} }.
\end{align}
Equation~(\ref{eq:Info-flow-0_general}) explicitly shows that the information flow vanishes in thermal equilibrium when DB holds, $\mathcal{T}_i a_{ji}=\mathcal{T}_j a_{ij}$, as already follows from its definition~(\ref{eq:Iflow}). Furthermore, it trivially vanishes if the cross-correlations nullify.
If $a_{01}\neq 0$, the information flow can be expressed as 
\begin{align}\label{eq:Info-flow-0}
{\dot{I}}_{\to 0}  
=&
\left(\underline{\underline{\Sigma}}^{-1}\right)_{01} \frac{\dot{ Q}_0}{\gamma_0 \,a_{01}\mathcal{T}_0}
=
\frac{- \langle X_1 X_0\rangle }{\langle X_1 X_0\rangle^2  -\langle X_1^2\rangle \langle X_0^2\rangle }\frac{\dot{ Q}_0}{\gamma_0 \,a_{01}\mathcal{T}_0}
%\nn =&
,
\end{align}
%The heat flow is shown in Fig.~\ref{fig:heat} and Fig.~\ref{fig:heat_n=1-2}.
% 
revealing that the information flow out of and into $X_0$ necessarily nullifies, if the heat flow is zero (if $a_{01}\neq 0$). 

%Figure~\ref{fig:info} shows the information flows for the case $n=1$.
The information flow is shown in Fig.~\ref{fig:heat} together with the heat flow. Along the unidirectional coupling axis $a_{01}=0$, there is net information flow from $X_0$ to $X_1$, but no net work applied to $X_0$ ($\dot{Q}_0=\dot{W}_0=0$). %The subsystem described by $X_0$ equilibrates. 
Thus, it is indeed sensible to consider $X_1$ a ``sensor'' and the coupling a {``sensing interaction''}. If the unidirectional coupling is reversed ($a_{10}=0$), the heat flow is always positive, $\dot{Q}_0>0$, i.e., an active swimmer eventually heats up its surrounding. In this case, there is as well a nonzero information flow, which is directed from the source of propulsion (e.g., the flagella) to the particle. This is also reasonable, as the propulsion force ``carries'' information: one could, on average, reconstruct the position of the flagella by only monitoring $X_0$. 

For non-reciprocal, bidirectional coupling, the information flow can be positive or negative, depending on whether the ``sensing'', or the ``active force'' is stronger.
It seems intuitive to consider $X_0$ a feedback-\textit{controlled} system, only if the net information flow out of $X_0$ is {positive}, i.e., the controller ``knows'' more about $X_0$ then vice versa. According to this definition, the \textit{control regime} is given if $|a_{10}|>|a_{01}|$ (blue regions in the middle panel of Fig.~\ref{fig:heat}). 
This is exactly the regime where we have detected the negative heat flow, i.e., here the controller may extract energy from a single heat bath (under isothermal conditions). Note that this observation is consistent with the generalized second law~(\ref{eq:GeneralizedSecondLaw}) which does not predict, but allow for a negative heat flow in this very regime only.

Interestingly, we find that another intriguing phenomenon may occur (only) when the information flow is negative, namely, the \textit{suppression of thermal fluctuations}. The latter can be measured by a reduced second moment $\langle X_0^2 \rangle < \langle X_0^2 \rangle_{a_{01}=0}$, which we have displayed in Fig.~\ref{fig:heat} (right panel). 
{In the blue areas, the second moment is reduced, thus, the feedback has the same effect as stiffening the trap. {This resembles the situation in a recent experiment involving colloids in an optical trap~\cite{Wallin2008}, where time-delayed feedback was used to effectively stiffen a trap.} Thermal fluctuation suppression can further be viewed as ``isothermal compression'' of a single-molecule gas, which represents, for example, an important step in the cycle of a (colloidal) heat engine~\cite{Martinez2016,Blickle2012}. It also implies noise-reduction, which is desired in various experimental setups, and indeed one of the main applications of feedback control~\cite{Steck2006,Cohadon1999,Vinante2008}. Interestingly, by only varying $a_{10}$ (which does not explicitly appear in the equation for $X_0$), one can vary between fluctuation enhancement (isothermal expansion), and fluctuation suppression (isothermal compression). The suppression of thermal fluctuations is limited to the area where one direction of the coupling is attractive ($a_{ij}<0$) while the revers direction is repulsive ($a_{ij}>0$). We find it quite remarkable that whenever ${\dot{I}}_{\to 0}<0$, such that $X_1$ can be viewed as a controller, it either yields a suppression of the fluctuations of $X_0$ (reduction of Shannon entropy), or a heat flow from the bath to $X_0$ (reduction of medium entropy).} % We note that also in the case $n=2$, which we turn to next, the regime of thermal fluctuation suppression (not shown here) is limited to the area of negative information flow, i.e, to the control regime.

Lastly, we detect a further counter-intuitive property appearing exclusively in non-reciprocal super-systems: \textit{there are nonequilibrium steady states, where all information flows nullify} (note that ${\dot{I}}_{\to 1}= -{\dot{I}}_{\to 0}$ for $n\,=\,1$). Thus, the subsystems may be driven out of equilibrium just due to their interaction (as signaled by finite dissipation), but without exchanging any information with each other. 

\subsection{Two non-reciprocal interactions, n=2}
For higher $n$, the explicit expressions for the information flow are quite cumbersome. For example, for $n=2$, 
%\begin{widetext}
	\begin{align}\label{eq:info_n=2}
	{\dot{I}}_{\to 0}
	&
	=\frac{-a_{00}}{\gamma_0}+
	\frac{
		\gamma_0^{-1}(a_{00}\langle X_0^2\rangle+a_{01}\langle X_0X_1\rangle+a_{02}\langle X_0X_2\rangle)(\langle X_1X_2\rangle^2-\langle X_1^2\rangle \langle X_2^2\rangle)	
	}
	{\langle X_0^2\rangle \langle X_1X_2\rangle^2+\langle X_1^2\rangle\langle X_0X_2\rangle^2+\langle X_2^2\rangle \langle X_0X_1\rangle^2 -2 \langle X_0X_1\rangle \langle X_0X_2\rangle \langle X_1X_2\rangle-\langle X_0^2\rangle \langle X_1^2\rangle \langle X_2^2\rangle}.
	%
%	\\
%	%
	\end{align}
Again, Eq.~(\ref{eq:info_n=2}) reflects that the existence of a nonzero information flow necessarily implies that the d.o.f. are cross-correlated among each other. 
However, different from the case $n=1$, there is \textit{no proportionality} between heat and information flow. % (this proportionality was previously reported in~\cite{allahverdyan2009thermodynamic}). 
In contrast, we find that for $n>1$, the relationship between those quantities becomes more complicated. 
To better understand their relationship, let us consider the cases $n=1,2$ again with the parameter setting from~(\ref{K_noise_II}, \ref{eq:Controller_n=1}), shown in Fig.~\ref{fig:heat_n=1-2}. 
Remarkably, despite the different nature of the super-system and the different type of memory, the information flow maps look almost \textit{identical} for $n=1$ and $2$. This indicates that the information flow is almost exclusively affected by the direct coupling (here from $X_0$ to $X_1$), which is, in principle, the same in both cases [given by the force $-(1/\tau)X_1$]. Thus, different from the energy flows, the information exchange is not affected by the additional indirect coupling though a third d.o.f. in the case $n=2$. 
Furthermore, we again find that the areas of negative heat flow [blue region in Fig.~\ref{fig:heat_n=1-2} (a)] appear in the control regime of $\dot{I}_{\to 0}<0$ [blue region in Fig.~\ref{fig:heat_n=1-2} (b)]. We note that, as in the case $n=1$, the regime of thermal fluctuation suppression (not shown here) is limited to the area of negative information flow, i.e, to the control regime.

Apart from these similarities, we observe a phenomenon which only occurs for $n>1$ and non-reciprocal coupling, that is, 
the existence of NESS where $X_0$ is out of equilibrium with broken FDR and ${\dot{I}}_{0\to} < 0$, but $\dot{Q}_0 = 0$. 
%This means that, counter-intuitively, \textit{a (sub-)system may be out of equilibrium due to  about its environment, without dissipating}. 
Considering the entropy balance~(\ref{eq:Sj1}), the entropy produced in $X_0$ due to the non-reciprocal coupling force, is transported only in the form of information. This state corresponds to the aforementioned non-Markovian NESS with zero dissipation (see Sec.~\ref{SEC:revHeatFlow}).
%
%
%
%%%%%%%%%
%
%
%
%%%%%%%%%%%%%%%%%%%%%%%%%%%%%%%%%%%%%%%%%
%
%
%
%
%%%%%%%%%%%%%%%%%%
%
%
%

% % % % % % % % % % % % % % % % % % % % % % % % % % % % % % % % % %
\section{\uppercase{Mapping non-reciprocity onto temperature gradients}}\label{sec:hiddenTemp}
%-- a hidden temperature gradient as underlying driving mechanism
%
In the course of this paper, we have demonstrated that non-reciprocal coupling introduces ``activity'', or more generally, intrinsic nonequilibrium. 
In contrast, there are several other recent publications which discuss (hidden) temperature gradients between reciprocally coupled stochastic d.o.f. as possible mechanisms that fuel active motion, see, e.g.,~\cite{Li2019,roldan2018arrow,netz2018fluctuation}. 
In this section we show that, in some cases, non-reciprocal coupled systems can indeed be mapped onto a reciprocally coupled system with an internal temperature gradient.

Consider the non-reciprocal system with $n=1$ and $a_{01}a_{10} \neq 0$, 
\begin{align} 
\begin{cases}\label{eq:ASymmetric-network}
\gamma_0 \dot{X}_0 &=a_{00}X_0 +a_{01}X_1 + \xi_0\\
\gamma_1 \dot{X}_1 &=a_{10}X_0 +a_{11}X_1 + \xi_1. 
\end{cases}
\end{align} 
We now introduce new variables 
$\sqrt{| a_{01} |}\,\widetilde{X}_0=X_0$, $\sqrt{| a_{10} |}\,\widetilde{X}_1=X_1$, and $|a_{01}|\widetilde{\mathcal{T}}_0 =\mathcal{T}_0$, $|a_{10}|\widetilde{\mathcal{T}}_1 =\mathcal{T}_1$. We note that if the $X_j$ are position-like d.o.f., their scaling should indeed be accompanied by scaling of the temperatures due to the connection between temperatures and the time-derivative of the positions. In this way, we find
%\begin{subequations}
\begin{align}
\begin{cases}\label{eq:Symmetric-network}
\gamma_0 \dot{\widetilde{X}}_0 &=  a_{00} \widetilde{X}_0 + \text{sgn}(a_{01})\sqrt{a_{01}a_{10} } \widetilde{X}_1 + \widetilde{ \xi}_0\\
\gamma_1 \dot{\widetilde{X}}_1 &= \text{sgn}(a_{10}) \sqrt{ a_{10}a_{01}  }  \widetilde{X}_0 +  a_{11} \widetilde{X}_1 + \widetilde{\xi}_1,
\end{cases}
\end{align}
with $\langle \widetilde{ \xi}_i(t)\widetilde{ \xi}_j(t) \rangle =2 k_\mathrm{B}\widetilde{\mathcal{T}}_j\gamma_j \delta_{ij}\delta(t-t')$.
%\end{subequations} $\mathcal{T}_1 =|a_{01}|\mathcal{T}' $
If $a_{01}a_{10}\!>\!0$, this system has \textit{reciprocal} coupling. Further, even if $\mathcal{T}_0=\mathcal{T}_1$, it involves a temperature gradient. The symmetric system~(\ref{eq:Symmetric-network}) could, for example, model the angles of two vanes in different heat baths, coupled by a torsion spring~\cite{Sekimoto2010}. % (see Fig.~\ref{fig:heat}). 

As well-known~{\cite{Parrondo1996,Sekimoto2010}}, such a reciprocally coupled system %[as well as the example~(\ref{eq:Symmetric-network_ex})]
equilibrates if, and only if, 
$\widetilde{\mathcal{T}}_1 = \widetilde{\mathcal{T}}_0 $ $\Leftrightarrow |a_{01}|{\mathcal{T}}_1= |a_{10}|\mathcal{T}_0 $. The equilibrium condition found in this way is identical to the equilibrium condition~(\ref{eq:Fulfill-DB}) found from DB and FDR.
Importantly, these considerations are not restricted to the case $n=1$. In Appendix~\ref{sec:Mapping}, we give an explicit example for a non-reciprocal system with $n=2$ that can be mapped onto a reciprocally coupled one, if $a_{ij}a_{ji} > 0,~\forall i,j\,\in\{0,1,2\}$. Again, this mapping yields the identical equilibrium conditions as~(\ref{eq:Fulfill-DB}), and the strategy can be generalized to larger $n$. Thus, here we have shown that, when the equilibrium model with non-reciprocal coupling and temperature difference is mapped onto a reciprocally coupled system —which is potentially realisable by a mechanical setup— the temperature difference vanishes.

Now we turn to the impact of this scaling on the thermodynamic quantities, using $n=1$ as an illustration. For the heat flows, we find the relations
\begin{align}\label{eq:MappingHeat}
%{\delta \widetilde{w}}_\mathrm{0}  &=  \text{sgn}(a_{01})\sqrt{a_{01}a_{10} }\, \widetilde{X}_1  \circ \mathrm{d}{\tilde{X}}_0 
%\nn
%&=  
%= X_1  \circ \,\mathrm{d}{X}_0 =   {\delta   w %}_\mathrm{0} /|a_{10}| ,
{\delta \widetilde{q}}_\mathrm{0}  &=   (\gamma_0 \dot{\widetilde{X}}_0-\widetilde{\xi}_0 ) \circ \mathrm{d}{\tilde{X}}_0 =  {\delta q }_\mathrm{0}  /|a_{01}|
\nn
{\delta \widetilde{q}}_\mathrm{1}  &=   (\gamma_1 \dot{\widetilde{X}}_1-\widetilde{\xi}_1 ) \circ \mathrm{d}{\tilde{X}}_1 =  {\delta q }_\mathrm{1}  /|a_{10}|
.
% 
%\nn
%{\mathrm{d} \tilde{u}}_\mathrm{0}/\mathrm{d}t  &=   a_{00} \circ \dot{\tilde{X}}_0 =  |a_{10}|\,{\delta  u }_\mathrm{0}/\mathrm{d}t .
\end{align} 
This further means
\begin{align}
\Delta {\tilde{s}}_\mathrm{tot} =&\, \Delta {\tilde{s}}_\mathrm{sh} +\frac{  {\delta \tilde{q}}_\mathrm{0} } {\widetilde{\mathcal{T}}_0}
+ \frac{{\delta \tilde{q}}_\mathrm{1} } {\widetilde{\mathcal{T}}_1}
=
\, \Delta {\tilde{s}}_\mathrm{sh} +\frac{  {\delta  {q}}_\mathrm{0} } {|a_{01}|\widetilde{\mathcal{T}}_0}
+ \frac{{\delta  {q}}_\mathrm{1} } {|a_{10}|\widetilde{\mathcal{T}}_1}
%\nn=&
= \Delta {s}_\mathrm{sh} +\frac{  {\delta {q}}_\mathrm{0} } {\mathcal{T}_0}
+ \frac{{\delta {q}}_\mathrm{1} } { \mathcal{T}_1}=\Delta {s}_\mathrm{tot},
\end{align}
i.e., the EP in the scaled model is \textit{identical} to the EP in the original model, while the energy flows in general differ.

We conclude that the two ``driving mechanisms'', that is, non-reciprocal coupling (with $a_{ij}a_{ji}>0$), or a temperature gradient, can formally not be distinguished on the level of EP. 
This mapping also builds a bridge to active matter models where temperature gradients between reciprocally coupled stochastic d.o.f. fuel the active motion~\cite{Li2019,roldan2018arrow,netz2018fluctuation}. 
It should be emphasized, however, that a scaling as employed here cannot be found if $a_{ij}a_{ji}\leq 0$ (which, interestingly, includes unidirectional coupling, e.g., the AOUP model). This suggests that non-reciprocal coupling is the more general way to introduce intrinsic non-equilibrium.

\section{Underdamped dynamics}\label{sec:underdampedCases}
So far, we have focused on overdamped descriptions, which are appropriate when the inertia is negligible, or if one is mainly interested in the dynamics above the ballistic timescale. However, in certain situations the inertia terms might yield contributions to thermodynamic quantities that are crucial to obtain a physically consistent description, even above the ballistic timescale. This is, e.g., the case for feedback systems with very short delay times~\cite{Loos2019}. More importantly in the present context, this is also true for Markovian systems that are simultaneously coupled to multiple heat baths at different temperatures, since then energy may be transferred between different heat baths via the kinetic energy of the system, see, e.g., \cite{Hondou2000}. 
	Therefore, we dedicate this last section to the consideration of inertia effects in the presence of non-reciprocal coupling. 
We will pay special attention to the following two aspects:
	(i) Does the equilibrium nature of the non-reciprocal models which fulfill~(\ref{eq:Fulfill-DB}) persist when we account for inertia terms? (ii) Is our calculation of the heat flow consistent with underdamped dynamics?

First, %we show the validity of the equipartition theorem for the non-reciprocal equilibrium systems. To this end, we will first 
we revisit the mapping from Sec.~\ref{sec:hiddenTemp}, now for underdamped dynamics.
To this end, we add the inertia terms to (\ref{eq:ASymmetric-network}), i.e., 
\begin{align} 
\begin{cases}\label{eq:underdampedNR}
m_0 \ddot{X}_0 +\gamma_0 \dot{X}_0 &=a_{00}X_0 +a_{01}X_1 + \xi_0\\
m_1 \ddot{X}_1 +\gamma_1 \dot{X}_1 &=a_{10}X_0 +a_{11}X_1 + \xi_1. 
\end{cases}
\end{align} 
Again, we introduce the variables 
$\sqrt{| a_{01} |}\,\widetilde{X}_0=X_0$, $\sqrt{| a_{10} |}\,\widetilde{X}_1=X_1$, and $|a_{01}|\widetilde{\mathcal{T}}_0 =\mathcal{T}_0$, $|a_{10}|\widetilde{\mathcal{T}}_1 =\mathcal{T}_1$,
and obtain
\begin{align}
\begin{cases}\label{eq:underdampedR}
m_0\ddot{\widetilde{X}}_0+ 	\gamma_0 \dot{\widetilde{X}}_0 &=  a_{00} \widetilde{X}_0 + \text{sgn}(a_{01})\sqrt{a_{01}a_{10} } \widetilde{X}_1 + \widetilde{ \xi}_0
\\
m_1\ddot{\widetilde{X}}_1+	\gamma_1 \dot{\widetilde{X}}_1 &= \text{sgn}(a_{10}) \sqrt{a_{01}a_{10}  }  \widetilde{X}_0 +  a_{11} \widetilde{X}_1 + \widetilde{\xi}_1.
\end{cases}
\end{align}
As before, the mapping yields a reciprocal system, if $a_{01}a_{10}>0$. %Again, the mapping can readily be generalized to $n>1$.

Due to the explicit inclusion of the inertia terms in the underdamped case, we can now consider the equipartition theorem, which represents yet another measure for equilibrium. If the reciprocal system~(\ref{eq:underdampedR}) is in equilibrium, traditional thermodynamics tells us that equipartition holds, thus
\begin{align}
\left \langle \frac{\widetilde p_0^2}{2m_0} \right \rangle = \left \langle \frac{k_\mathrm{B}\widetilde{\mathcal{T}}_0}{2} \right \rangle , ~~ 
\left \langle \frac{\widetilde p_1^2}{2m_1} \right \rangle = \left \langle \frac{k_\mathrm{B}\widetilde{\mathcal{T}}_1}{2} \right \rangle ,
\end{align}  with $p_{0,1}=m_{0,1} \dot{\widetilde{X}}_{0,1}$, and that $\widetilde{\mathcal{T}}_0=\widetilde{\mathcal{T}}_1$. Transforming back to the original variables, this corresponds to
\begin{align}
\left \langle \frac{ p_0^2}{2 |a_{01}| m_0} \right \rangle = \left \langle \frac{k_\mathrm{B}{\mathcal{T}}_0}{2  |a_{01}| } \right \rangle , ~~ 
\left \langle \frac{\widetilde p_1^2}{2 |a_{10}| m_1} \right \rangle = \left \langle \frac{k_\mathrm{B}{\mathcal{T}}_1}{2|a_{10}| } \right \rangle .
\end{align} 
Hence, also 
the non-reciprocal (underdamped) system~(\ref{eq:underdampedNR}) fulfills the equipartition theorem if $\widetilde{\mathcal{T}}_0=\widetilde{\mathcal{T}}_1 \Rightarrow a_{10}{\mathcal{T}}_0=a_{01}{\mathcal{T}}_1 $. This condition is in agreement with the equilibrium condition from DB for overdamped dynamics, Eq.~(\ref{eq:Fulfill-DB}). We emphasize that the arguments presented here [including the mapping (\ref{eq:underdampedR})] can readily be generalized to $n>1$.

Next, we consider the heat flow in the presence of inertia. To this end, we consider as a specific example the case $m_{0,1}=m$,  $\gamma_{0,1}=\gamma$, $a_{00}=a_{11}=- \text{sgn}(a_{10}) \sqrt{ a_{10}a_{01}  }$, and introduce the new variable $\tilde \kappa = \text{sgn}(a_{10}) \sqrt{ a_{10}a_{01}  }$, to simplify the notation. Then, (\ref{eq:underdampedR}) reduces to
\begin{align}
\begin{cases}
m \ddot{\widetilde{X}}_0+ 	\gamma \dot{\widetilde{X}}_0 &=  -\tilde \kappa ( \widetilde{X}_0 -  \widetilde{X}_1 )+ \widetilde{ \xi}_0
\\
m \ddot{\widetilde{X}}_1+	\gamma \dot{\widetilde{X}}_1 &=- \tilde \kappa ( \widetilde{X}_1  - \tilde{X}_0) + \widetilde{\xi}_1.
\end{cases}
\end{align}
For this system, the heat flow between system $\widetilde X_0$ and its bath has been calculated in Ref.~\cite{Parrondo1996} [see there Eq.~(A16), and note the different sign convention]. In our notation, it reads
% \tilde \kappa =   \kappa
% \gamma = \lambda
\begin{align}
\dot{\widetilde Q}_0 = - k_\mathrm{B}
 \frac{   \tilde \kappa  }{2    (\gamma + \tilde \kappa m/\gamma) }(\widetilde{ \mathcal{T}}_0-\widetilde{ \mathcal{T}}_1).
\end{align}
Transforming back to the original variables [and recalling (\ref{eq:MappingHeat}) which also holds in the underdamped description], this yields the heat flow (recall $a_{01}a_{10}>0$)
\begin{align}\label{eq:underdampedHeat}
\dot{Q}_0 =-
\frac{ |a_{01}| k_\mathrm{B}  \,\text{sgn}(a_{10}) \sqrt{ a_{01}a_{10}   } }{2    (\gamma + \text{sgn}(a_{10}) \sqrt{ a_{01}a_{10}   } m/\gamma) }\left(\frac{ \mathcal{T}_0}{|a_{01}|}-\frac{\mathcal{T}_1}{|a_{10}|}\right)
=
\frac{-  k_\mathrm{B}|a_{01}|   }{2    (\gamma + \text{sgn}(a_{10}) \sqrt{ a_{10}a_{01}  } m/\gamma) }\left(\frac{ a_{10} \mathcal{T}_0-a_{01} \mathcal{T}_1}{\sqrt{ a_{01}a_{10}  } }\right)
.
\end{align}
Thus, the heat flow vanishes if (\ref{eq:Fulfill-DB}) is fulfilled. This exactly agrees with the condition that the heat flow in the overdamped description vanishes.

Finally, we take the overdamped limit $m/\gamma \to 0$ of (\ref{eq:underdampedHeat}), which yields
\begin{align}
\lim_{(m/\gamma) \to 0} \dot{Q}_0 =
-  k_\mathrm{B} 
\frac{  |a_{01}|  }{2   \gamma  }\left(\frac{ a_{10} \mathcal{T}_0-a_{01} \mathcal{T}_1}{\sqrt{ a_{10}a_{01}  } }\right)
.
\end{align}
This is indeed identical to (\ref{eq:Q0_n=1}) for the given parameters, confirming the consistency of our considerations.
%
%%%%%%%%%%%%%%%%%%%%%%%%%%%%%%%%%%%%%%%%%%%%%%%%%%%%%%%%%%%%%%%%%
%%%%%%%%%%%%%%%%%%%%%%%%%%%%%%%%%%%%%%%%%%
%%%%%%%%%%%%%%%%%%%%%%%%%%%%%%%%%%%%%%        CONCL       %%%%%%%%%%%%%%%%%%%%%%%%%%%%%%%%%%%%%%%%%%%%%%%%%%%%%%%%%
%%%%%%%%%%%%%%%%%%%%%%%%%%%%%%%%%%%%%%%%%%%%%%%%%%%%%%%%%%%%%%%%%%%%%%%%%%%%%%%%%%%%%%%%%%%%%%%%%%%%%%%%%%%
\section{Conclusion \label{sec:Conclusion}}
This paper addresses the thermodynamic implications of non-reciprocal coupling between stochastic d.o.f., which is a form of non-conservative interaction appearing in various artificial or natural complex systems across the fields. The most important result is that the occurrence of a non-reciprocal coupling alone implies nonequilibrium, as indicated by a broken detailed balance and fluctuation-dissipation relation, and is automatically associated with a net energy and information flow. 
Remarkably, we found that under special conditions (specifically if $a_{ij}\mathcal{T}_j=a_{ji}\mathcal{T}_i$), non-reciprocal system can reach a state of thermal equilibrium, despite begin simultaneously coupled to two heat baths at different temperatures. To prove the equilibrium nature of this state, we have considered a variety of equilibrium measures, that is, the fluctuation-dissipation relation, detailed balance, the equipartition theorem when we additionally include inertia terms, zero total entropy production, zero heat and information flows. In these equilibrium situations, the non-reciprocal system with internal temperature gradient can be formally mapped onto a reciprocal one at isothermal conditions, giving a mathematical explanation for the observed exceptions.
 Another key result is that a non-reciprocal coupling between isothermal d.o.f. may induce, for one of the two d.o.f., a negative heat flow (while the total dissipated energy is always positive), meaning that energy is extracted from the bath. This shows a crucial difference between the thermodynamic implications of a non-conservative (non-reciprocal) interaction vs. a non-conservative external force, which could only induce a positive heat flow (as dictated by the second law). 
 	Both, the existence of isothermal systems with negative heat flow, and the existence of thermal equilibrium despite temperature gradients, are intriguing phenomena, which significantly depart from the thermodynamic behavior of reciprocal systems.
 	Indeed, giving intuitive explanations appears to be challenging. We hope that this manuscript will stimulate fruitful discussions and future research on this matter.

As different, exemplary representatives of non-reciprocal systems, we have considered active matter or feedback-controlled systems. 
While a single unidirectional coupling makes $X_1$ a ``propulsion mechanism'' and $X_0$ an ``active swimmer'', a single non-reciprocal \textit{bi}directional coupling may make $X_1$ a ``feedback controller'' that operates on $X_0$. Moreover, when the controller knows more about the controlled system than vice versa (indicated by an information flow to the controller), some major goals of feedback control can be achieved, including thermal fluctuation suppression, and energy extraction of the heat bath (i.e., a {negative} heat flow) making $X_1$ a minimal version of a continuously operating ``Maxwell demon''. The latter can only be achieved if (i) the information flow is directed from the system to the controller and (ii) the controller applies negative feedback, i.e., a feedback force pointing away from the delayed position of $X_0$. 

Whereas one non-reciprocal coupling ($n=1$) only induces exponentially decaying memory in the corresponding non-Markovian equation for the single d.o.f. (e.g., $X_0$), the interplay of multiple linear non-reciprocal interactions ($n>1$) allows to generate non-monotonic memory, which, in turn, is typical for time-delayed feedback control. From a thermodynamic point of view, the cases $n=1$ and $n=2$ share the main characteristics. However, there is indeed a crucial difference, that is, the heat and information flows are not proportional to each other, if $n>1$. Thus, one can find for $n=2$ some interesting nonequilibrium steady states %Along the way, we report several further interesting types of NESS, 
which only occur for $n>1$ and non-reciprocal coupling. On the one hand, mutually coupled systems can be driven out of equilibrium due to their interaction, without at the same time exchanging any information. On the other hand, for a different non-reciprocal coupling topology, one can also find a state where one of these subsystems is in a NESS where it exports the entropy exclusively in the form of information without displaying a heat flow (no entropy is exported to the bath). 

We close this paper by giving some perspectives on future research. 

In our present paper, we have shown that, under certain conditions, non-reciprocal forces can be mapped onto temperature gradients. Moreover, it is known that 
non-reciprocal couplings may result from gradients of chemical potentials~\cite{Saha2020,Saha2019, You2020}. This is, e.g., the case in the cellular sensor model~\cite{hartich2016sensory}, used as an example in this paper (Sec.~\ref{SEC:Model}).
Thus, it seems worth to systematically explore in the future whether, and under which conditions, a mapping onto other thermodynamic ``forces'' is feasible. 
A major focus of recent research is the search of meaningful thermodynamic descriptions for active systems. This is indeed not the topic of the present work, and we have here merely scratched the surface of this issue. For example, it is generally not possible to access the full dissipation of a complex living system, as long as not all underlying bio-chemical processes are fully known, understood, and also \textit{observable}. 
The last point, i.e., the observability is related to another main problem in this context, that is, the thermodynamic treatment of auxiliary, or effective variables, which lack of a clear physical interpretation, as it is the case for the variable $X_j$ in the AOUP model. As we have pointed out several times throughout the paper, in such a situation the meaning of, e.g., the total EP is questionable. 
To account for this fact, we have discussed the different measures of (non)equilibrium on the Markovian and non-Markovian level of description. However, the detailed balance condition or the fluctuation-dissipation relation only yield a binary classification (equilibrium or not), but cannot quantify the distance from equilibrium. Finding out an appropriate way to do this is discussed, e.g., in \cite{Li2019}.
An interesting line of research in the context of observability and auxiliary variables, is the search of ``effective thermodynamic'' descriptions~\cite{Herpich2020,Polettini2017,Pietzonka2019}. For a similar underdamped model with $n=1$, different ways to obtain an ``effective thermodynamic'' description, were recently compared in~\cite{Herpich2020}. A generalization towards higher $n$ (and overdamped models) represents a nontrivial but certainly worthwhile direction for future research. It would also be interesting to investigate the here observed special types of NESS, e.g., with zero dissipation but nonzero information flow, from this perspective.

Also, regarding the different measures for (non)equilibrium, our preliminary observations indicate that for $n>1$, there are non-reciprocal systems that fulfill FDR but violate DB, i.e., are nonequilbrium models with fluctuation-dissipation relations. It might be interesting to study the corresponding information flows for these cases.

%Going a bit further away from the system considered here, another main
In this paper, we have analyzed the thermodynamic properties of small stochastic systems of few colloids with non-reciprocal couplings. As a next step, one could think about the implications of our findings for larger systems with numerous non-reciprocal couplings, which are, as a matter of fact, already realized in recent experiments~\cite{Lavergne2019}. Indeed, the non-reciprocity is found to yield intriguing clustering collective behavior. %
At this point, we also aim to note that in non-linear dynamics and network science, studying the effects of symmetry-broken coupling on the collective behavior is already a well-established research field~\cite{Loos2016}. For example, the existence of chimera states, a special type of clustering, was linked to symmetry-broken coupling~\cite{Premalatha2015}, and shown to persist in the presence of discrete delay~\cite{Zakharova2016} and Gamma-distributed memory~\cite{kyrychko2013amplitude}.

Lastly, the unidirectionally coupled ring system studied here is very similar to the reservoir computers investigated in~\cite{larger2017high,li2018deep}. A reservoir computer of this type may be experimentally realized by a laser network~\cite{rohm2019reservoir,rohm2018multiplexed}, or by coupled RC circuits~\cite{kish2012electrical,snider2011minimum}. Another link to machine learning is the similarity between the unidirectional ring and recurrent neural networks~\cite{Xu2004}, used for example for reinforcement learning. In these contexts, the connection between non-reciprocal coupling and information flow discussed here might be of particular importance. 
Noteworthy, the architecture of the unidirectional ring considered here also resembles the architecture of a Brownian clock~\cite{barato2016cost}, which, in contrast, has discrete dynamics.
%
%

% % % % % % % % % % % % % % % % % % % % % % % % % % % % % % % % % % % % % % % % % % % % % % % % % % % % % % % % % % % % % % % % % % % % % % % % % %
% % % % % % % % % % % % % % % % % % % % % % % % % % % % % % % % % % % % % % % % % % % % % % % % % % % % % % % % % % % % % % % % % % % % % % % % % %
% % % % % % % % % % % % % % % % % % % % % % % % % % % % % % % % % % % % % % % % % % % % % % % % % % % % % % % % % % % % % % % % % % % % % % % % % %
%%%%%%%%%%%%%%%%%%%%%%%%%%%%%
\begin{acknowledgments}
This work was funded by the Deutsche Forschungsgemeinschaft (DFG, German Research Foundation) - Projektnummer 163436311 - SFB 910. 
\end{acknowledgments}
%\end{document}
% % % % % % % % % % % % % % % % % % % % %
% % % % % % % % % % % % % % % % % % % % %
%%%%%%%%%%%%%%%%%%%%%%%%%%%%%%%%%%%%%%%%%%%%%%%%%%%%%%%%%%%%%%%%%
%%%%%%%%%%%%%%%%%%%%%%%      APPENDIX       %%%%%%%%%%%%%%%%%%%%%
\appendix
%
% % % % % % % % % % % % % % % % % % % % % % % % % % % % % % % %
\section{\uppercase{Memory kernel for up to three coupled systems}}\label{sec:MemoryRing} %\label{sec:MemoryRing}
Here we derive the memory kernel for the case $n=1$ and $n=2$ by projecting the equations for $X_{j>0}$ onto $X_0$. To this end, we solve the equations for $X_{j\in \{1,2\}} $ in frequency space, making use of their linearity (we want to emphasize that their linearity is irrespective of the question whether the equation of $X_0$ is linear, thus, is result also applies to cases with nonlinear $f_0$). First, we apply the Laplace transformation
 $\mathcal{L}[X_j(t)](s)=\int_{0}^{\infty} X_j(t)e^{-st}\mathrm{d}s$
to the LE $ \gamma_j \dot{X}_j= \sum_{l=0}^1 a_{j l} X_l +\xi_{j},$
which yields
\begin{align}\label{eq:Laplace_Xj_allG01}
-X_j(0) \gamma_j + s \gamma_j {\hat{X}_j}(s)= & \sum_{l=0}^n  a_{j l} \hat {X}_l(s) +  \,\hat{\xi}_j(s)  .
\end{align}
Since we are interested in steady-state dynamics in this paper, we can safely set $X_j(0)\equiv 0$ without loss of generality. We therewith obtain for all $j$,
\begin{align}\label{eq:Laplace_Xj_allG}
 {\hat{X}_j}(s)= & \sum_{l \neq j }^n  \frac{ a_{j l} \hat {X}_l(s)}{(s \gamma_j - a_{j j})}  + \frac{1}{(s \gamma_j - a_{j j})} \,\hat{\xi}_j(s)  .
\end{align}

Let us first consider the case $n=1$. We plug (\ref{eq:Laplace_Xj_allG}) for $j=1$ into the equation~(\ref{eq:Laplace_Xj_allG01}) for 
$X_0$ and immediately find
\begin{align} 
 s \gamma_0 {\hat{X}_0} = & a_{00} \hat {X}_0  + \frac{a_{01}a_{1 0}}{(s \gamma_1 - a_{1 1})}  \hat {X}_0  + \frac{a_{01}}{(s \gamma_1 - a_{1 1})}\,\hat{\xi}_1 +  \,\hat{\xi}_0 .
\end{align}
Now we make use of the convolution theorem and the linearity of the Laplace transformation to
transform back to real space, obtaining the non-Markovian process~(\ref{eq:LE-X0}) with a memory kernel given by the inverse Laplace transformation of $\frac{a_{01}a_{1 0}}{(s \gamma_1 - a_{1 1})} $, as explicitly given in~(\ref{eq:ExampleI}). Analogously, one finds the Gaussian colored noise in~(\ref{eq:LE-X0})
\begin{align}
\nu(t)=& \int_{0}^{t}  \frac{a_{01}  }{\gamma_{1}} e^{a_{11}(t-t')/\gamma_1}  \, \xi_{1}(t')  \mathrm{d}t',
\end{align}
with correlation $C_\nu(\Delta t) = \langle \nu(t)\nu(t+\Delta t)\rangle$
\begin{align}
C_\nu(\Delta t) =& \frac{a_{01}^2  }{\gamma_{1}^2} \int_{0}^{t}\int_{0}^{t+\Delta t}  e^{a_{11}/\gamma_1[(t-t')+(t-t'')+\Delta t]}  \,\langle \xi_{1}(t')\xi_{1}(t'') \rangle  \mathrm{d}t'\mathrm{d}t'' %\nn
=     \frac{ k_\mathrm{B} \mathcal{T}_1 a_{01}^2  }{a_{11}} \,e^{a_{11}\Delta t/\gamma_1}   [1- e^{a_{11}2t/\gamma_1}].
\end{align}
In the steady state ($t \to \infty$), the second term vanishes (if $a_{11}<0$), yielding the correlation from~(\ref{eq:ExampleI}).

Next, we derive the memory kernel for the case $n=2$. 
yields
\begin{align} 
{\hat{X}_1} = & \frac{    a_{1 0} \hat {X}_0 }{(s \gamma_1 - a_{1 1})}    +   \frac{a_{1 2}[ a_{2 0} \hat {X}_0  + a_{2 1} {\hat {X}_2}   ]}{(s \gamma_2 - a_{2 2})(s \gamma_1 - a_{1 1})}     + \mathcal{O}(\hat{\xi}_1) + \mathcal{O}(\hat{\xi}_1 \hat{\xi}_2), \\
{\hat{X}_2} = & \frac{a_{2 0} \hat {X}_0 }{(s \gamma_2 - a_{2 2})}   + \frac{a_{2 1}  [ a_{1 0} \hat {X}_0   + a_{1 2} {\hat {X}_1}   ]  }{(s \gamma_2 - a_{2 2})(s \gamma_1 - a_{1 1})}   + \mathcal{O}(\hat{\xi}_2) + \mathcal{O}(\hat{\xi}_1\hat{\xi}_2).  
\end{align}
%....
%\
%
Note that because $\langle \xi_1 \xi_2 \rangle =0$, the terms $\mathcal{O}(\hat{\xi}_1\hat{\xi}_2)$ will not contribute in the end [see, e.g., Eq.~(\ref{eq:colorednoise_1})] and can thus be neglected. We can further simplify the expressions to
\begin{align} \label{eq:Laplace_Xj_allG1}
 \hat {X}_1 
= & \left[ \frac{    a_{1 0}  (s \gamma_2 - a_{2 2}) + a_{1 2} a_{2 0} }{(s \gamma_2 - a_{2 2})(s \gamma_1 - a_{1 1}) - a_{1 2}  a_{2 1}  }     \right]  \hat {X}_0
  + \mathcal{O}(\hat{\xi}_1) , 
  \\
  \hat{X}_2   
= & \left[  \frac{a_{2 0}(s \gamma_1 - a_{1 1}) +a_{2 1}  a_{1 0}   }{(s \gamma_2 - a_{2 2})(s \gamma_1 - a_{1 1})- a_{2 1}    a_{1 2}  }  
 \right] \hat {X}_0
+ \mathcal{O}(\hat{\xi}_2) .  
\end{align}
Substituting~(\ref{eq:Laplace_Xj_allG1}) in~(\ref{eq:Laplace_Xj_allG01}) for $j=0$, one obtains
\begin{align} \label{EQ:KernelLaplaceSpace_n=2}
 \gamma_0 s{\hat{X}_0} %= &    a_{0 0}  \hat {X}_0  + a_{0 1} \hat {X}_1   + a_{0 2} \hat {X}_2     + \mathcal{O}(\hat{\xi}_0). \nn
    = &  ~  a_{0 0}  \hat {X}_0  + \left[ \frac{    a_{0 1} a_{1 0}  (s \gamma_2 - a_{2 2}) + a_{0 1}a_{1 2} a_{2 0} }{(s \gamma_2 - a_{2 2})(s \gamma_1 - a_{1 1}) - a_{1 2}  a_{2 1}  }    +  \frac{a_{0 2}a_{2 0}(s \gamma_1 - a_{1 1}) +a_{0 2}a_{2 1}  a_{1 0}   }{(s \gamma_2 - a_{2 2})(s \gamma_1 - a_{1 1})- a_{2 1}    a_{1 2}  }  
 \right] \hat {X}_0 + {\sum_{j=0}^2}\mathcal{O}(\hat{\xi}_j) \nn
 = & ~ a_{0 0}  \hat {X}_0(s)  + \hat{K}(s) \hat {X}_0(s)     + {\sum_{j=0}^2}\mathcal{O}(\hat{\xi}_j).
\end{align}
% % % % % % % % % % % % %
Finally, transforming back to real space yields the non-Markovian process~(\ref{eq:LE-X0}) with a memory kernel given by the inverse Laplace transformation of $\hat{K}(s)$.

Specifically, for the unidirectionally coupled ring system with $n=2$ illustrated in Fig.~\ref{fig:kernels} (a), which is described by the set of equations~(\ref{eq:LE-X0}) with $a_{jj} =-(p+\kappa)$, $a_{j j+1}= p$, $a_{j+1} =\kappa$, and $\gamma_j=1$ for $j \in \{0,1,2\}$, Eq.~(\ref{EQ:KernelLaplaceSpace_n=2}) simplifies to 
\begin{align}
 s{\hat{X}}_0 =&  
  \left[ \frac{ p^3+\kappa^3 +2 p \kappa (s+p+\kappa)  }{  (s+p+\kappa)^2- p \kappa  }    
   \right]  \hat{X}_0  
 - \kappa \hat{X}_0 - p \hat{X}_0 +{\sum_{j=0}^2}\mathcal{O}(\hat{\xi}_j).
 %
% p \left[ \frac{ p^2 }{ (s+p+\kappa)^2- p \kappa  } + \frac{ \kappa (s+p+\kappa)}{ (s+p+\kappa)^2- p \kappa  }  \right]  \hat{X}_0   
%  +\kappa  \left[ \frac{ \kappa^2 }{ (s+p+\kappa)^2- p \kappa  } + \right. \left. \frac{ p (s+p+\kappa)}{ (s+p+\kappa)^2- p \kappa  } \right]  \hat{X}_0  
%  \nn
%  &
%  - \kappa \hat{X}_0 - p \hat{X}_0 +\mathcal{O}(\hat{\xi}_j).
\end{align}
In real space, this memory kernel (given in square brackets) reads~(\ref{eg:kernel_n=3}).

% % % % % % % % % % % % % % % % % % %
%%%%%%%%%%%%%%%%%%%%%%%%%%%%%%%%%%%%%%%%%%%%%%%%%%%%%%%%%%%%%%%%%%
%
\section{\uppercase{Memory kernel and noise correlations}}\label{sec:deriveKandNu}
We here derive the memory kernel and colored noise in the model~(\ref{K_noise_II}), i.e., a unidirectional ring with $n=2$. Analogously to the derivation in Appendix \ref{sec:MemoryRing}, we first apply the Laplace transformation 
to the LE 
$\dot{X_j}(t)=(1/{\tau}) \left[ X_{j-1}(t)- X_{j}(t) \right] +   \,\xi_j(t)/\gamma_1$ for $j\in \{1,2\}$, and set $X_j(0)\equiv 0$, obtaining
\begin{align}\label{eq:Laplace_Xj_all}
{\hat{X}_j}(s) %=& \frac{\tau^{-1}\hat{X}_{j-1}(s)  + \hat{\xi}_j(s)/\gamma'}{ s+\tau^{-1}}
 =  \frac{\tau^{-1}\hat{X}_{j-1}(s)   }{ s+\tau^{-1}} +  \frac{ \gamma_1^{-1} \hat{\xi}_j(s) }{ s+\tau^{-1}}.
\end{align}
Iteratively substituting the solution~(\ref{eq:Laplace_Xj_all}) for $j=1$ into (\ref{eq:Laplace_Xj_all}) for $j=2=n$, yields
\begin{align}\label{eq:Laplace_Xn}
{\hat{X}_2}(s)  =  \frac{\tau^{-2}   }{( s+\tau^{-1})^2}\hat{X}_{0}(s)   +  \frac{ \tau^{-1}}{( s+\tau^{-1})^2}  \frac{\hat{\xi}_1(s) }{\gamma_1}  +  \frac{ 1 }{ s+\tau^{-1}} \frac{\hat{\xi}_2(s) }{\gamma_1}
.
\end{align}
Now we transform back to the real space via inverse Laplace transformation. In~(\ref{eq:Laplace_Xn}) we identify the Laplace-transform of the Gamma-distribution
${\displaystyle \mathcal{L}\left[K_{j}(t)\right]( s)={\frac {\tau^{-j}}{[s+\tau^{-1}]^{j}}}} $ with the Gamma-distributed kernels $K_{j}(t)={\frac {t^{j-1}}{\tau^{j}(j-1)!}}\, e^{- t /\tau}$, and find
\begin{align}\label{eq:RealSpace_Xn}
{X}_2(t)
=& \int_{0}^{t} K_{2}(t-t') {X}_{0}(t') \,\mathrm{d}t' + \nu_2(t)
\end{align}
with the Gaussian colored noise
\begin{align}
\nu_2(t)=& \int_{0}^{t} \frac{\tau}{\gamma_1} K_{2}(t-t')  \, \xi_{1}(t') + \frac{\tau}{\gamma_1} K_{1}(t-t')  \, \xi_{2}(t') \,\mathrm{d}t'. 
%\nu_n(t)=&\sum_{j=1}^{n}\int_{0}^{t} \frac{\tau}{\gamma' n} K_{j}(t-t')  \, \xi_{n-j+1}(t') \,\mathrm{d}t'.  %\sqrt{\frac{\tau \epsilon_{n-j+1}}{n} }
\end{align}
Replacing $X_2(t)$ from~(\ref{eq:RealSpace_Xn}) in the Markovian LE $\gamma \dot{X}_0(t) = a_{00}X_0(t) +k X_2(t) +\xi_0(t)$, yields the non-Markovian LE 
$$\gamma \dot{X}_0(t) = a_{00}X_0 +\int_{0}^{t} K(t-t') {X}_{0}(t') \,\mathrm{d}t' + \xi_0(t)+ \nu(t)$$
with the memory kernel $K(T) = k K_2(T) =  \frac {k}{\tau^{2}}T\, e^{- T /\tau}$ [as given in (\ref{K_noise_II})] and the colored noise $ \nu(t) =k \nu_2(t) $.

The noise correlations $C_{\nu}(\Delta t)=\langle \nu(t)\nu(t+\Delta t)\rangle$ can be calculated exactly for an arbitrary $\Delta t>0$, as we will show in the following. First, we use the properties of the white noise, e.g, $\langle \xi_{1}(t')\xi_{2}(t'')\rangle = 0$, $\langle \xi_{j}(t')\xi_{j}(t'')\rangle = 2 k_\mathrm{B} \mathcal{T}_1 \gamma_1 \delta(t'-t'')$, and integrate out the {Dirac} delta distributions, yielding % $\langle \xi_{n-j+1}(t')\xi_{n-i+1}(t'') \rangle=2\gamma' k_\mathrm{B}\mathcal{T}'\delta_{i,j}\delta(t'-t'')$
\begin{align}\label{eq:colorednoise_1}
C_{\nu}(\Delta t) %= k^2 \langle \nu_2(t)\nu_2(t+\Delta t)\rangle 
=&
=
 \frac{\tau^2 k^2 2 k_\mathrm{B} \mathcal{T}_1  }{\gamma_1 } \int_{0}^{t} \sum_{j=1,2}  K_{j}(t-t') K_{j}(t-t'+\Delta t)   \mathrm{d}t'  .
\end{align}
Plugging in,
$K_{1}(T)=\tau^{-1}\, e^{- T /\tau}$
and
$K_{2}(T)=\tau^{-2} T\, e^{- T/\tau}$, this can be further be simplified to

\begin{align}
C_\nu(\Delta t) =&~~
%\sum_{j=1,2}  \frac{\tau k^2}{\gamma_1} \int_{0}^{t}  K_{j}(t-t') K_{j}(t-t'+\Delta t)   \mathrm{d}t' \nn=&
  \frac{  k^2 2 k_\mathrm{B} \mathcal{T}_1   }{\gamma_1   } e^{- \Delta t /\tau} \int_{0}^{t}  [1 + \tau^{-2}(t-t')(t-t'+\Delta t)]\, e^{- 2(t-t') /\tau}   \,  \mathrm{d}t' 
\nn
\stackrel{u=2(t-t')/\tau}{=}&~~
% \frac{\tau k^2}{\gamma_1 \tau^{2}} e^{- \Delta t /\tau} \int_{0}^{2 t/\tau}  [1 +   (u /2) (u /2+\Delta t /\tau )]\, e^{-  u }   \,  \mathrm{d}u
%\nn
% =&
 \frac{\tau}{2} \frac{ k^2\, 2 k_\mathrm{B} \mathcal{T}_1   }{\gamma_1  } e^{- \Delta t /\tau} \int_{0}^{2 t/\tau}  [1+  (u/2)\,(\Delta t/\tau) +  u^2/4 ]\, e^{-  u }   \,  \mathrm{d}u .
\end{align}
As we are interested in steady states, we now take the limit $t \to \infty$ and then perform the integration using
$%\begin{align}
\int_{0}^{\infty} x^p e^{-x}\mathrm{d}x =p!,
$ %\end{align}
which readily yields the noise correlation given in~(\ref{K_noise_II}).
We note that the transient correlation could be calculated similarly by instead using the incomplete Gamma function.

\section{Analytical solutions \label{sec:analytical_solutions}}
As indicated by~(\ref{eq:Q0_general},\,\ref{eq:ShannonSubTotal},\,\ref{eq:Stot_general},\,\ref{eq:generalFormula_Infoflow},\,\ref{eq:Info-flow-0}), various (thermo-)dynamic quantities can be calculated on the basis of the correlations $\langle X_iX_{j} \rangle $. Further, the steady-state pdf $\rho_{n+1}$ %[see Eq. (\ref{eq:FPE-extended})] 
is, due to the linearity of the model, a Gaussian-distribution with zero mean and the covariance matrix $(\underline{\underline{\Sigma}})_{i j}=\langle X_i X_j\rangle$. Thus, it is fully determined by all the correlations $\langle X_i X_j\rangle$.

Here we sketch how analytical expressions for these correlations can be obtained
for arbitrary system sizes $n$.
To this end, we transform Eqs.~(\ref{eq:Network}) via the Fourier transformation
$\tilde{X}_j(s)=\int_{-\infty}^{\infty} X_j(t)e^{-i \omega t}\mathrm{d}t$, which readily yields %
\begin{align}\label{eq:Laplace_Xjb}
i \omega \underline{\underline{\gamma}}  {\underline{\tilde{X}}}(\omega)=& \underline{\underline{a}} {\underline{\tilde{X}}}(\omega) + {\underline{\tilde{\xi}}(\omega) }
%\\
\Rightarrow {\underline{\tilde{X}}}(\omega)=  \underbrace{\left(  i \omega \underline{\underline{\gamma}} -\underline{\underline{a}}\right)^{-1} }_{=\underline{\underline{\tilde{\lambda}}}(\omega)
} \,{\underline{\tilde{\xi}}(\omega) },
\end{align} 
with the Green's function in Fourier-space $\underline{\underline{\tilde{\lambda}}}(\omega)$, determined by the inverse of the topology matrix $\underline{\underline{a}}$.
Using the well-known relationship between spatial correlations and the Green's function from linear response theory~\cite{Hanggi1982}
\begin{align}
C(\Delta t) = \frac{D_0}{\pi}\int_{-\infty}^{\infty}\tilde{\lambda}(-\omega)\tilde{\lambda}(-\omega)\,e^{-i\omega \Delta t}\mathrm{d}\omega,
\end{align}
one readily finds
\begin{align}\label{eq:Solution-Corr}
\langle X_j^2\rangle & =  \sum_{p=0}^{n} \frac{k_\mathrm{B}\mathcal{T}_p\gamma_p}{\pi} \int_{-\infty}^{\infty} \tilde{\lambda }_{jp}(\omega) \tilde{\lambda }_{jp} (-\omega) \,\mathrm{d}\omega,
\\
\langle X_jX_{l}\rangle & =  \sum_{p=0}^{n} \frac{k_\mathrm{B}\mathcal{T}_p \gamma_p}{\pi} \int_{-\infty}^{\infty} \tilde{\lambda }_{jp}(\omega) \tilde{\lambda }_{l p} (-\omega) \,\mathrm{d}\omega.
\end{align}
These are analytical expressions for all correlations for arbitrary system sizes $n$. %To evaluate them, different tricks can be used. We aim to note, that t

While this strategy in principle yields analytical expressions for various (linear) systems (which can, e.g., be
numerically integrated),
explicit closed-form solutions are only available for specific cases, where the inverse Fourier transformation is known (see~\cite{Geiss2019, Mackey1994} for some explicit results). For example, the correlations for $n=1$ read~\cite{Crisanti2012}
\begin{align}\label{eq:Corr_n=1}
\underline{\underline{\Sigma}} =
\begin{pmatrix}
\langle X_0^2\rangle &  \langle X_0X_1\rangle\\
\langle X_1X_0\rangle  &\langle X_1^2\rangle  \\
\end{pmatrix}
=
k_\mathrm{B}
\begin{pmatrix}
\frac{\mathcal{T}_1 a_{01}^2-\mathcal{T}_0 a_{01}a_{10}+\mathcal{T}_0 a_{11} (a_{00}+a_{11}\gamma_0/\gamma_1)}{(a_{00}+a_{11}\gamma_0/\gamma_1)(a_{01}a_{10}-a_{00}a_{11})}  &   \frac{-\mathcal{T}_1 a_{00}a_{01}-\mathcal{T}_0 a_{11}a_{10}\gamma_0/\gamma_1}{(a_{00}+a_{11}\gamma_0/\gamma_1)(a_{01}a_{10}-a_{00}a_{11})} \\
\frac{-\mathcal{T}_1 a_{00}a_{01}-\mathcal{T}_0 a_{11}a_{10}\gamma_0/\gamma_1}{(a_{00}+a_{11}\gamma_0/\gamma_1)(a_{01}a_{10}-a_{00}a_{11})}   & \frac{\mathcal{T}_0 a_{10}^2\gamma_0/\gamma_1-\mathcal{T}_1 a_{01}a_{10}\gamma_0/\gamma_1+\mathcal{T}_1 a_{00} (a_{00}+a_{11}\gamma_0/\gamma_1)}{(a_{00}+a_{11}\gamma_0/\gamma_1)(a_{01}a_{10}-a_{00}a_{11})} \\
\end{pmatrix}
%\\
%%
%\langle X_0^2\rangle &= \frac{T_1 a_{01}^2-T_0 a_{01}a_{10}+T_0 a_{11} (a_{00}+a_{11})}{(a_{00}+a_{11})(a_{01}a_{10}-a_{00}a_{11})}
%%
%,~~
%\langle X_0X_1\rangle &= \frac{-T_1 a_{00}a_{01}-T_0 a_{11}a_{10}}{(a_{00}+a_{11})(a_{01}a_{10}-a_{00}a_{11})}
%%
%\\
%\langle X_1X_0\rangle &= \frac{-T_1 a_{00}a_{01}-T_0 a_{11}a_{10}}{(a_{00}+a_{11})(a_{01}a_{10}-a_{00}a_{11})}
%%
%\\
%%
%\langle X_1^2\rangle &= \frac{T_0 a_{10}^2-T_1 a_{01}a_{10}+T_1 a_{00} (a_{00}+a_{11})}{(a_{00}+a_{11})(a_{01}a_{10}-a_{00}a_{11})}
.
\end{align}
We
could not find general closed-from solutions for the problem with $n>1$.

The matrix inversion is indeed possible up to very large system sizes,if the coupling is sparse (e.g., for unidirectionally coupled ring systems). To evaluate the integrals, the residue theorem can be used. However, this requires finding the roots of a polynomial of order $n+1$. Using computer algebra systems, this can be done reasonably fast up to about $n=10$. We also note, for the case $\mathcal{T}_{j>0}=0$, solutions up to $n\sim 10^4$ can be found in this way. 
\section{\uppercase{On the terminology of positive and negative feedback}}\label{APP:Feedback}
In control theory, it is common to characterize feedback loops as positive or negative feedback, according to the question whether the 
force points towards, or away from the desired state once there is a perturbation from it, see, e.g.,~\cite{Bechhoefer2005}. In the following, we check of which type the feedback considered in this paper is.
We recall that in the present case, the control problem, which is given in (\ref{eq:LE-X0_Feedback}) or, equivalently, in (\ref{eq:LE-X0}), reads
$$ \gamma_0 \dot{X}_0(t ) = a_{00}X_0 (t) +F_\mathrm{c} + f_0 +\mu(t),$$
with the feedback force $F_\mathrm{c}= 
\int K(t-t') X_0(t')\mathrm{d}t'$. For the sake of illustration, let us explicitly consider the limit $n \to \infty$, where the notation simplifies while the following reasoning is the same for any $n$. Thus, $F_\mathrm{c}= 
k X_0(t-\tau)$. This control problem can be alternatively expressed as
$$ \gamma_0 \dot{X}_0(t ) = (a_{00}+k)X_0 (t) -k[X_0(t)-X_0(t-\tau)] + f_0 +\mu(t),$$
suiting to the picture of a colloidal in a static harmonic trap of stiffness $a_{00}+k < 0$, and subject to a co-moving feedback trap centered around $X_0(t-\tau)$ and with stiffness $k$. When $X_0(t) \equiv X_0(t-\tau)$, the control term $-k[X_0(t)-X_0(t-\tau)]$ vanishes, thus, this is a ``non-invasive'' control. In contrast, when the system is perturbed from the delayed state, it may yield a positive or negative force on $X_0$. Specifically, if $k>0$, we have a negative force $-k[X_0(t)-X_0(t-\tau)] < 0$ whenever $X_0(t) > X_0(t-\tau)$. On the other hand, this force is positive, whenever $X_0(t) < X_0(t-\tau)$.

{Now, it is clear that the feedback with $k>0$ is pointing \textit{towards} the past state, $X_0(t-\tau)$. Therefore, this case is denoted positive feedback. On the contrary, the feedback force is always pointing \textit{away} from $X_0(t-\tau)$ if $k<0$.}
\section{\uppercase{Mapping onto a reciprocal super-system}\label{sec:Mapping}}
% % % % % %%%%%%%%%%%%%%%%%%%%%%%%%%%%%%%%%%%%%
%Let us now also briefly consider cases with $n>1$, in particular, the case $n=2$. 
In Sec.~\ref{sec:hiddenTemp} we discuss the mapping of an non-reciprocal coupled system onto a reciprocal system (with different temperatures) for systems with $n=1$. In this Appendix, we generalize this idea to larger system sizes.

%We recall that DB is fulfilled if 
%$..$. Can we find an analogous mapping 
%As before, we cannot find such a mapping for general cases, but only under certain conditions. For example, we suspect that $a_{01}a_{01}a_{02}a_{20}a_{12}a_{21}\neq 0$, and $a_{01}a_{10} > 0$,  $a_{02}a_{20} > 0$, and $a_{12}a_{21} > 0$, are necessary conditions (in agreement with the DB condition). Among the systems that respect these conditions, a 
A specific type of non-reciprocal coupling topology, for which we could find a mapping, is 
\begin{align} \label{eq:ASymmetric-network-n=2}
\begin{pmatrix}
\gamma_0 \dot{{X}}_0   \\
\gamma_1  \dot{{X}}_1  \\
\gamma_2 \dot{{X}}_2\\
\end{pmatrix}
=
\begin{pmatrix}
a_{00}  &  r  &  v  \\
p   & a_{11}  & v  \\
p & r  & a_{22} \\
\end{pmatrix}
\underline{\dot{{X}}}
\underline{{{\xi}}}
,
\end{align} 
i.e., the two outward connections of each sub-system are identical (e.g., the coupling from $X_0$ to $X_1$ and from $X_0$ to $X_2$).
Networks of type~(\ref{eq:ASymmetric-network-n=2}) can be mapped onto a reciprocally coupled system via the coordinate transformation
$\widetilde{X}_0=\,X_0/\sqrt{| p |}$, $\widetilde{X}_1=\,X_1/\sqrt{| r |}$,
$\widetilde{X}_2=\,X_2/\sqrt{| v |}$,
and $\widetilde{\mathcal{T}}_0 =\mathcal{T}_0/|p|$, $\widetilde{\mathcal{T}}_1 =\mathcal{T}_1/|r|$, $\widetilde{\mathcal{T}}_2 =\mathcal{T}_2/|v|$.
The corresponding reciprocal super-system reads
\begin{align}
\begin{pmatrix}
\gamma_0 \dot{\widetilde{X}}_0   \\
\gamma_1  \dot{\widetilde{X}}_1  \\
\gamma_2 \dot{\widetilde{X}}_2\\
\end{pmatrix}
=
\begin{pmatrix}
a_{00}  &  \text{sgn}(r)\sqrt{ r p }   &  \text{sgn}(v)\sqrt{v p }  \\
\text{sgn}(r)\sqrt{ r p }   & a_{11}  &  \text{sgn}(r)\sqrt{r v  }   \\
\text{sgn}(v)\sqrt{v p } & \text{sgn}(r)\sqrt{r v  }  & a_{22} \\
\end{pmatrix}
\underline{\dot{\widetilde{X}}}
\underline{{\widetilde{\xi}}}
,
\end{align}
with $\langle \widetilde{ \xi}_i(t)\widetilde{ \xi}_j(t) \rangle =2 k_\mathrm{B}\widetilde{\mathcal{T}}_j\gamma_j \delta_{ij}\delta(t-t')$.
As in the case $n=1$, we cannot find such a mapping for general cases, but only under certain conditions, specifically:
$a_{ij}a_{ji} > 0,~\forall i,j~\in\{0,1,2\}$.
 As in the case $n=1$, we use the following argument: a reciprocally (i.e., ``mechanical") system equilibrates in the absence of temperature gradients, i.e., $\widetilde{\mathcal{T}}_0=\widetilde{\mathcal{T}}_1=\widetilde{\mathcal{T}}_2$, which in the original coordinates gives the same condition (\ref{eq:Fulfill-DB}) as we found from DB.
 %\begin{align}
 %a_{ij} \mathcal{T}_j =a_{ji} \mathcal{T}_i. 
 %\end{align}
The mapping presented in this Appendix can straightforwardly be generalized to arbitrary $n\in \mathbb{N}$.
%
%
%
%
%
% % % % % % % % % % % % % % % % % % % % % %
\section{\uppercase{Mutual information}}\label{sec:AppInfo}
Here we discuss the relation in steady states, between the information flow considered in Sec.~\ref{sec:Info}, and the multivariate generalization of the mutual information given in (\ref{def:mutualInfo}).
We
start with considering the total derivative of $\mathcal{I}$ from Eq.~(\ref{def:mutualInfo}), that is,
\begin{align}
\mathcal{\dot{I}}& = 
 \int \!  \underbrace{\partial_t \rho_{n+1}(\underline{x})}_{  = (*)  }   \ln \frac{  \rho_{n+1}(\underline{x})}{ \rho_{1}(x_0)...\rho_{1}(x_n) }  \,\mathrm{d}\underline{x} 
+ \int \!  \rho_{n+1}(\underline{x})  %\times\nn& 
 \left\{  -\frac{  \partial_t  \rho_{n+1}(\underline{x})}{ \rho_{n+1}(\underline{x})} \right.\left. - \frac{  \partial_t [\rho_{1}(x_0)\rho_{1}(x_1)...\rho_{1}(x_n) ] }{ \rho_{1}(x_0)\rho_{1}(x_1)...\rho_{1}(x_n) }   \right\} \mathrm{d}\underline{x}  .
\end{align}
We substitute $(*)$ by utilizing the {multivariate} FPE~(\ref{eq:FPE-vector}) $\partial_t \rho_{n+1} = - \sum_{j=0}^n \partial_{x_j} J_j$, and find
\begin{align}
\mathcal{\dot{I}} 
=&
\sum_{j=0}^n\int \partial_{x_j} J_{j}(\underline{x})   \ln \frac{ \rho_{1}(x_0)...\rho_{1}(x_n) } {  \rho_{n+1}(\underline{x})} \mathrm{d}\underline{x} 
- \int \underbrace{ \partial_t  \rho_{n+1}}_{\to 0} \mathrm{d}\underline{x} 
%\nn&
- \int   \frac{\rho_{n+1} }{ \rho_{1}(x_0)...\rho_{1}(x_n)  }  \underbrace{\partial_t[\rho_{1}(x_0)...\rho_{1}(x_n)]}_{\to 0}  \mathrm{d}\underline{x}  
\nn=&
%=
\sum_{j=0}^n\int \partial_{x_j} J_{j}(\underline{x})   \ln \frac{ \rho_{1}(x_0)...\rho_{1}(x_n) } {  \rho_{n+1}(\underline{x})} \mathrm{d}\underline{x} 
.
\end{align}
Let us now consider the individual summands. By application of basic properties of the logarithm and the natural boundary conditions, we find
%, and integrate out as many d.o.f. as possible via partial integration. For example, if $j\neq 1$, we perform the following step
%

\begin{align}
\int \!  \partial_{x_j} J_j(\underline{x}) \ln\frac{ \rho_{1}(x_0)\rho_{1}(x_1)..\rho_{1}(x_n) }{  \rho_{n+1}(\underline{x})}   \,\mathrm{d}\underline{x}
%=&... =
=&  \iint \! \ln \frac{ \rho_{1}(x_j) \rho_{1} (x_{i\neq j}) }{  \rho_{n+1}(\underline{x})}  \partial_{x_j} J_j  \,\mathrm{d}\underline{x} 
%-\int \! \ln \rho_{1}(x_{i})   \partial_{x_j} J_j   \,\mathrm{d}\underline{x} 
-\iint \! \ln \rho_{1}(x_{i }) \underbrace{\left[  J_j \right]_{-\infty}^{\infty}}_{\to 0 } \,\mathrm{d}\underline{x}_{\neq j} 
\nn= &
\iint \! \ln \frac{\rho_{1}(x_j)}{  \rho_{n+1}(\underline{x})}  \partial_{x_j} J_j  \,\mathrm{d}\underline{x}
=
{\dot{I}}_{\to j} 
\end{align}
%\end{align}
Thus, the change of mutual information is given by the sum over all information flows, $\sum_{j=0}^n {\dot{I}}_{\to j}  =\mathcal{\dot{I}}$. (As was shown in~\cite{allahverdyan2009thermodynamic}, the information flow ${\dot{I}}_{\to j}$ is actually the ``time-shifted mutual information'' with the time shift applied to $X_j$.) 
%On the other hand, as can easily be seen from its definition~(\ref{def:mutualInfo}), $\mathcal{{I}}$ is a conserved quantity in steady-states, $\mathcal{\dot{I}}=0$.
%
%
%
%%%%%%%%%%%%%%%%%%%%%%%%%%%%%%%%%%%%%%%%%%%%%%%%%%%%%%%%%%%%%%%%%
\bibliography{AA_references.bib}
%%%%%%%%%%%%%%%%%%%%%%%%%%%%%%%%%%%%%%%%%%%%%%%%%%%%%%%%%%%%%%%%
%%%%%%%%%%%%%%%%%%%%%%%%%%%%%%%%%%%%%%%%%%%%%%%%%%%%%%%%%%%%%%%%%
%%%%%%%%%%%%%%%%%%%%%%%%%%%%%%%%%%%%%%%%%%%%%%%%%%%%%%%%%%%%%%%%%
%%%%%%%%%%%%%%%%%%%%%%
% % % % % % % % % % % % % % % % % % % % % % % % % % % % % % % % % % % % % % % % % % % % % % % % % % % % % % % % % % % % % % % % % % % % % % % % % % % % %
%%%%%%%%%%%%%%%%%%%%%%%%%%%%%%%%%%%%%%%%%%%%%%%%%%%%%%%%%%%%%%%%%%
%%%%%%%%%%%%%%%%%%%%%%%%%%%%%%%%%%%%%%%%%%%%%%%%%%%%%%%%%%%%%%%%%%%%%%%%%%%%%%%%%%%%%
%%%%%%%%%%%%%%%%%%%%%%%%%%%%%%%%%%%%%%%%%%%%%%%%%%%%%%%%%%%%%%%%%%%%%%%%%%%%%%%%%%%%%%%%%%%%%%%%%%%%%%%%%%%
%%%%%%%%%%%%%%%%%%%%%%%%%%%%%%%%%%%%%%%%%%%%%%%%%%%%%%%%%%%%%%%%%%%%%%%%%%%%%%%%%%%%%%%%%%%%%%%%%%%%%%%%%%%
%%%%%%%%%%%%%%%%-----------ENDE-------%%%%%%%%%%%%%%%%%%%%%%%%%%%%%%%%%%%%%%%%%%%%%%%%%%%%%%%%
%%%%%%%%%%%%%%%%%%%%%%%%%%%%%%%%%%%%%%%%%%%%%%%%%%%%%%%%%%%%%%%%%%%%%%%%%%%%%%%%%%%%%%%%%%%%%%%%%%%%%%%%%%%
%%%%%%%%%%%%%%%%%%%%%%%%%%%%%%%%%%%%%%%%%%%%%%%%%%%%%%%%%%%%%%%%%%%%%%%%%
\end{document}